\documentclass[a4paper, 11pt]{article}
\pdfoutput=1

\usepackage{latexsym, amsmath, amsfonts, amssymb}
\usepackage{tikz-cd}
\usepackage[framemethod=TikZ]{mdframed}
\usetikzlibrary{decorations.pathmorphing, cd, decorations.markings, calc, patterns, shapes.symbols}
\usetikzlibrary{decorations.pathreplacing, calligraphy}
\usetikzlibrary{arrows}
\usetikzlibrary{shapes}
\usetikzlibrary{matrix}
\usetikzlibrary{positioning}
\usetikzlibrary{shapes.multipart}
\usepackage{diagbox}
\usetikzlibrary{positioning}
\usetikzlibrary{calc}
\usepackage{xstring}

\usepackage{mathrsfs}
\usepackage[american]{babel}
\usepackage{graphicx}
\usepackage{bbm}
\usepackage{cite}

\usepackage[colorlinks=true, citecolor=blue!90!black, linkcolor=blue!90!black, linktocpage=true, urlcolor=red!70!black]{hyperref}

\renewcommand{\baselinestretch}{1.2}
\newcommand{\wh}{\widehat}

\newcommand{\matht}[1]{{\ensuremath{\boldsymbol{#1}}}}
\newcommand{\tps}[2]{\texorpdfstring{#1}{#2}}

\newcommand{\eg}{\textit{e.g.}}    

\newcommand{\ie}{\textit{i.e.}}

\numberwithin{equation}{section}

\newcommand{\be}{\begin{equation}} \newcommand{\ee}{\end{equation}}
\newcommand{\bea}{\begin{equation} \begin{aligned}} \newcommand{\eea}{\end{aligned} \end{equation}}

\newcommand{\cA}{\mathcal{A}}
\newcommand{\cB}{\mathcal{B}}
\newcommand{\cC}{\mathcal{C}}

\newcommand{\cF}{\mathcal{F}}

\newcommand{\cM}{\mathcal{M}}

\newcommand{\cW}{\mathcal{W}}

\newcommand{\bB}{\mathbb{B}}

\newcommand{\bZ}{\mathbb{Z}}

\def\su{\mathfrak{su}}

\def\repa{\raise4pt\hbox{$\square$}\mkern-14mu\raise-4pt\hbox{$\square$}}
\def\repab{\overline{\raise4pt\hbox{$\square$}\mkern-14mu\raise-4pt\hbox{$\square$}\mkern-1mu}}

\DeclareMathOperator{\Tr}{Tr}

\DeclareMathOperator{\Pexp}{Pexp}

\begin{document}
\thispagestyle{empty}
	\fontsize{12pt}{20pt}
	\begin{flushright}
		SISSA  05/2024/FISI
	\end{flushright}
	\vspace{13mm}  
	\begin{center}
 
  {\huge On the Symmetry TFT of \\ \vskip 13pt
  Yang--Mills--Chern--Simons theory
  }
  \bigskip

		{\large Riccardo Argurio$^a$, \, Francesco Benini$^{b , \, c , \, d}$, \, Matteo Bertolini$^{b , \, d}$ \\[3mm]\, Giovanni Galati$^{a}$, \, Pierluigi Niro$^{e}$
  
				\bigskip
				
				{\it
					$^a$ Physique Th\'eorique et Math\'ematique and International Solvay Institutes\\
Universit\'e Libre de Bruxelles, C.P. 231, 1050 Brussels, Belgium  \\[.2em]
					$^b$ SISSA, Via Bonomea 265, 34136 Trieste, Italy \\[.2em]
					$^c$ ICTP, Strada Costiera 11, 34151 Trieste, Italy \\[.2em]
					$^d$ INFN, Sezione di Trieste, Via Valerio 2, 34127 Trieste, Italy \\[.2em]
                        			$^e$ Mani L. Bhaumik Institute for Theoretical Physics, Department of Physics and Astronomy, University of California, Los Angeles, CA 90095, USA \\ [.2em]
					}}
		
	\end{center}

\bigskip 

\begin{abstract}
\vskip 5pt
\noindent 
Three-dimensional Yang--Mills--Chern--Simons theory has the peculiar property that its one-form symmetry defects have nontrivial braiding, namely they are charged under the same symmetry they generate, which is then anomalous. This poses a few puzzles in describing the corresponding Symmetry TFT in a four-dimensional bulk. First, the braiding between lines at the boundary seems to be ill-defined when such lines are pulled into the bulk. Second, the Symmetry TFT appears to be too trivial to allow for topological boundary conditions encoding all the different global variants. We show that both of these puzzles can be solved by including endable (tubular) surfaces in the class of bulk topological operators one has to consider. In this way, we are able to reproduce all global variants of the theory, with their symmetries and their anomalies. We check the validity of our proposal also against a top-down holographic realization of the same class of theories.
\end{abstract}

\newpage
\pagenumbering{arabic}
\setcounter{page}{1}
\setcounter{footnote}{0}
\renewcommand{\thefootnote}{\arabic{footnote}}

{\renewcommand{\baselinestretch}{.88} \parskip=0pt
	\setcounter{tocdepth}{2}

\tableofcontents}


\section{Introduction}
\label{sec: intro}

Our understanding of symmetries in Quantum Field Theory (QFT) has seen a substantial leap forward after the realization that any extended topological operator should be regarded as a symmetry element \cite{Gaiotto:2014kfa}. This new paradigm includes the standard definition of symmetry in QFT, but also considerably extends it. One convenient way to describe such generalized symmetries, partially inspired by the anomaly inflow mechanism that describes 't~Hooft anomalies \cite{Wess:1971yu, Callan:1984sa}, is through the so-called Symmetry Topological Field Theory (Symmetry TFT for short) \cite{Gaiotto:2014kfa, Ji:2019eqo, Gaiotto:2020iye, Apruzzi:2021nmk, Freed:2022qnc}. While (conventional) 't~Hooft anomalies are described through inflow by symmetry protected topological (SPT) phases, also known as invertible TQFTs \cite{Freed:2014iua} --- which essentially are classical topological field theories of the background fields in one higher dimension --- the Symmetry TFT is a full fledged (\ie{} possibly non-invertible) TQFT. As it turns out, such a topological theory is capable of encoding all Renormalization Group (RG) invariant data related to the symmetry of a QFT: the structure of the symmetry, its representations \cite{Lin:2022dhv, Bhardwaj:2023ayw, Bartsch:2023wvv}, 't~Hooft anomalies \cite{Kaidi:2023maf, Antinucci:2023ezl, Cordova:2023bja}, the set of ``global variants'' of the QFT obtained by topologically gauging subsets of the symmetry \cite{Gaiotto:2020iye}, the classification of possible gapped phases \cite{Bhardwaj:2023fca}, and so on. 

The Symmetry TFT has been mostly studied for the case of finite symmetries, but recently proposals for the extension to continuous symmetries have been put forward as well\cite{Brennan:2024fgj, Antinucci:2024zjp, Bonetti:2024cjk, Apruzzi:2024htg}. Other works on the Symmetry TFT include \cite{Kaidi:2022cpf, Zhang:2023wlu, Bhardwaj:2023wzd, Chen:2023qnv, Bashmakov:2023kwo, Sun:2023xxv, Baume:2023kkf, Huang:2023pyk, Bhardwaj:2023bbf, Bhardwaj:2024qrf}. Symmetry-TFT-like constructions in the context of string theory and holography include \cite{Hofman:2017vwr, Bergman:2020ifi, Benini:2022hzx, Apruzzi:2022rei, Bergman:2022otk, Heckman:2022muc, vanBeest:2022fss, Antinucci:2022vyk, Bashmakov:2022uek, Antinucci:2022cdi, Bah:2023ymy, Apruzzi:2023uma, Cvetic:2023pgm, Yu:2023nyn, Gould:2023wgl, DelZotto:2024tae}.

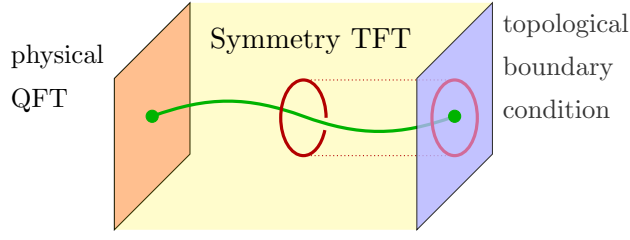
\begin{figure}[t]
\centering
\begin{tikzpicture}
	\draw [fill = red!30!white] (0,0) -- (1,1) -- (1,3) -- (0,2) node[left, align = left] {\small physical \\ \small QFT} -- cycle;
	\fill [fill = yellow, opacity = 0.2] (0,0) -- (4,0) -- (4,2) -- (5,3) -- (1,3) -- (0,2) -- cycle;
	\draw [very thick, green!70!black] (0.5, 1.5) to [out = 20, in = 160] (2.5, 1.5) to [out = -20, in = -160] (4.5, 1.5);
	\draw [densely dotted, red!70!black] (2.5, 1.98) to (4.5, 1.98);
	\draw [densely dotted, red!70!black] (2.5, 0.98) to (4.5, 0.98);
	\draw [fill = blue!30!white, opacity = 0.8] (4,0) -- (5,1) -- (5,3) node[below right, align = left] {\small topological \\ \small boundary \\ \small condition} -- (4,2) -- cycle;
	\filldraw [green!70!black] (0.5, 1.5) circle [radius = 0.08];
	\filldraw [green!70!black] (4.5, 1.5) circle [radius = 0.08];
	\draw [very thick, red!70!black] (2.8, 1.48) arc [start angle = 0, end angle = 341, y radius = 0.5, x radius = 0.3];	
	\draw [very thick, blue!30!white!60!red] (4.8, 1.48) arc [start angle = 0, end angle = 360, y radius = 0.5, x radius = 0.3];
	\node at (2.6, 2.5) {Symmetry TFT};
\end{tikzpicture}
\caption{\label{fig: slab}%
The prototype of a Symmetry TFT.}
\end{figure}

The description of the symmetry of a $d$-dimensional QFT works through a ``sandwich'' construction. The Symmetry TFT is placed on a $(d+1)$-dimensional slab between two copies of the $d$-dimensional spacetime manifold, pictured in Fig.~\ref{fig: slab}. On one boundary, referred to as the ``physical'' boundary, it is coupled to the QFT$_d$ of interest. On the other boundary, referred to as the ``topological'' boundary, one imposes a topological boundary condition. On general grounds, one expects the set of topological boundary conditions to be in one-to-one correspondence with the global variants of the QFT$_d$. The common lore is that specifying a boundary condition corresponds to prescribing (in a consistent way) which bulk operators of the Symmetry TFT can end on the topological boundary (green operator in Fig.~\ref{fig: slab}), and thus are trivialized at that boundary. Those operators constitute the set of charges (or representations) of the symmetry. On the contrary, the bulk operators which can be pushed to the boundary (red operator in Fig.~\ref{fig: slab}), modulo the ones that are trivialized there, constitute the set of topological symmetry defects of the QFT$_d$. In this way, from the fusion and the braiding between bulk operators, given a boundary condition, one is able to reconstruct the structure of the symmetry of the QFT$_d$ and the possible charges carried by its dynamical (non-topological) operators.

This construction, even in the simple case when the Symmetry TFT is Abelian, raises a few puzzles. If the symmetry defects of the QFT$_d$ braid among themselves, how can this be captured in the bulk of the Symmetry TFT that has one higher dimension? Given two extended operators that braid in $d$ dimensions, they do not braid in $d+1$ dimensions. A paradigmatic example is that of Yang--Mills--Chern--Simons (YM-CS) theories in $d=3$ dimensions. They typically have some finite 1-form symmetry whose symmetry defects are topological lines that braid among themselves.%
\footnote{The braiding phases, that depend on the Chern--Simons level, are the manifestation of an 't~Hooft anomaly for the 1-form symmetry.}
If those lines originate from lines of a putative 4d Symmetry TFT, how can the braiding be preserved in the bulk?

A natural way to construct the Symmetry TFT --- at least when the symmetry is invertible --- is to start with the QFT$_d$ coupled to background fields that extend in one higher dimension, as well as to the anomaly inflow Symmetry Protected Topological (SPT) phase,%
\footnote{An SPT phase is an invertible unitary TQFT, namely a TQFT with a one-dimensional Hilbert space on any closed spatial manifold \cite{Freed:2014iua}. Its Euclidean partition function is a phase.}
and then to gauge the symmetry in $d+1$ dimensions, namely to make the background fields dynamical. This in general produces a nontrivial TQFT$_{d+1}$.%
\footnote{For a generic non-invertible (categorical) symmetry $\cC$ in two  dimensions, the three-dimensional Symmetry TFT corresponds to the TFT whose symmetry structure is the Drinfeld center of $\cC$ (see \eg{} \cite{etingof2016tensor}). This TFT can be obtained from the Turaev--Viro construction \cite{turaev1992state}, which gives an explicit formulation of its partition function in terms of a state-sum formula (see \eg{} \cite{Thorngren:2019iar} for a more physically-oriented discussion). For three-dimensional QFTs with a symmetry described by a Modular Tensor Category (MTC), the Turaev--Viro construction can be generalized to give rise to a $4$d TQFT known as Crane--Yetter TQFT \cite{Crane:1993if, Crane:1994ji, Levin:2004mi} (see also \cite{douglas2018fusion, Barkeshli:2016mew, Bulmash:2020flp, Tata:2021jwp, Walker:2021esr} for further generalizations).}
When applied to YM-CS theories, that procedure yields a 4d TQFT that seems to lack the very line operators that braid on the boundary. In some extreme cases (\eg{} in $SU(N)_k$ with $N,k$ coprime) the bulk TQFT is still an SPT after gauging, and thus, naively, it does not have any topological operator. This is related to a second puzzle: an invertible TQFT has a single state on any spatial manifold and thus, if we regard the topological boundary conditions as states on the boundary manifold, it should have a unique boundary condition. On the other hand, the corresponding YM-CS theories can have multiple global variants (seemingly breaking the one-to-one correspondence with boundary conditions).

\begin{figure}[t]
\centering
\begin{tikzpicture}[scale=1.7]

\begin{scope}[shift={(-3.5, 0.1)}]
	\fill [white!70!green, opacity = 0.7] (-0.794, 0.6825) to ++(20: 0.7) to[out = 20, in = 20, looseness = 2] ++(-70: 0.6) to ++(-160: 0.59);  
	\draw [line width = 0.4] (-0.794, 0.6825) to ++(20: 0.7) to[out = 20, in = 20, looseness = 2] ++(-70: 0.6) to ++(-160: 0.04);  
	\draw [line width = 0.2, fill = orange] (0.18, 0.55) circle [radius = 0.02];  
	\fill [white!70!green, opacity = 0.7] (-0.178, -.081) to ++(20: 0.4) to[out = 20, in = 20, looseness = 2] ++(110:0.6) to ++(-160: 0.35);  
	\draw [line width = 0.4] (-0.178, -.081) to ++(20: 0.4) to[out = 20, in = 20, looseness = 2] ++(110:0.6) to ++(-160: 0.35);  
	\fill [white!90!green] (1.0, 0.67) +(-1.5, -0.5) arc (310: 670: 0.3);   
	\fill [white!90!green] (1.044, 0.937) +(-1.5, -0.5) arc (130: 490: 0.3);   
	\draw [line width = 0.4, white!90!green!60!black] (-0.49, 0.155) to ++(20: 0.5);  
	\draw [line width = 0.8, color = red] (1.0, 0.67) +(-1.5, -0.5) arc (310: 645: 0.3);  
	\draw [line width = 0.8, color = red] (1.044, 0.937) +(-1.5, -0.5) arc (130: 465: 0.3);  
	
	\node at (0.72, 0.55) {$=$};
\end{scope}

\begin{scope}[shift={(-1.5, 0.1)}]
	\filldraw [white!70!green, opacity = 0.7] (0.215, 1.0455) arc (107: -69.1: 0.3);   
	\filldraw [white!70!green, opacity = 0.7, shift={(0.4265, -0.1938)}] (0.215, 1.0455) arc (107: -69.1: 0.3);   

	\fill [white!70!green, opacity = 0.7] (0.215, 1.0455) -- (-0.794, 0.6825) -- (-0.605, 0.1128) -- (0.41, 0.4778) -- cycle;   
	\draw [line width = 0.4, shorten < = 33] (-0.605, 0.1128) -- (0.41, 0.4778);   
	\fill [white!70!green, opacity = 0.7, shift={(0.4265, -0.1938)}] (0.215, 1.0455) -- (-0.794, 0.6825) -- (-0.605, 0.1128) -- (0.41, 0.4778) -- cycle;   

	\fill [white!90!green] (1.0, 0.67) +(-1.5, -0.5) arc (310: 670: 0.3);   
	\fill [white!90!green] (1.044, 0.937) +(-1.5, -0.5) arc (130: 490: 0.3);   

	\draw [line width = 0.4] (0.215, 1.0455) -- (-0.794, 0.6825);   
	\draw [line width = 0.4, shift={(0.4265, -0.1938)}, shorten > = 2] (0.215, 1.0455) -- (-0.794, 0.6825);   
	\draw [line width = 0.4, white!90!green!60!black, shorten < = 6, shorten > = 22] (-0.605, 0.1128) -- (0.41, 0.4778);   
	\draw [line width = 0.4, shift={(0.4265, -0.1938)}] (-0.605, 0.1128) -- (0.41, 0.4778);   

	\draw [line width = 0.8, color = red] (1.0, 0.67) +(-1.5, -0.5) arc (310: 645: 0.3);
	\draw [line width = 0.8, color = red] (1.044, 0.937) +(-1.5, -0.5) arc (130: 465: 0.3);

	\node at (1.3, 0.55) {$=$};
\end{scope}

\begin{scope}[shift={(1.1, 0.1)}]
	\draw [fill = white!82!blue]  (0.3, 0.5) -- (1.8, -0.5) -- (1.8, 1) -- (0.3, 2) -- cycle;   

	\fill [white!73!green] (1.0, 0.67) arc (310: 670: 0.3);   
	\fill [white!73!green] (1.044, 0.937) arc (130: 490: 0.3);   
	\draw [line width = 0.5, densely dashed, dash phase = 2pt] (1.0, 0.67) arc (310: 645: 0.3);   
	\draw [line width = 0.5, densely dashed, dash phase = 1.8pt] (1.044, 0.937) arc (130: 465: 0.3);   

	\fill [white!70!green, opacity = 0.7] (0.715, 1.1855) -- (-0.794, 0.6825) -- (-0.605, 0.1128) -- (0.91, 0.6178) -- cycle;   
	\draw [line width = 0.4, shorten < = 33] (-0.605, 0.1128) -- (0.91, 0.6178);   
	\fill [white!70!green, opacity = 0.7, shift={(0.4265, -0.1938)}] (0.715, 1.1855) -- (-0.794, 0.6825) -- (-0.605, 0.1128) -- (0.91, 0.6178) -- cycle;   

	\fill [white!90!green] (1.0, 0.67) +(-1.5, -0.5) arc (310: 670: 0.3);   
	\fill [white!90!green] (1.044, 0.937) +(-1.5, -0.5) arc (130: 490: 0.3);   

	\draw [line width = 0.4] (0.715, 1.1855) -- (-0.794, 0.6825);   
	\draw [line width = 0.4, shift={(0.4265, -0.1938)}, shorten > = 2] (0.715, 1.1855) -- (-0.794, 0.6825);   
	\draw [line width = 0.4, white!90!green!60!black, shorten < = 6, shorten > = 48] (-0.605, 0.1128) -- (0.91, 0.6178);   
	\draw [line width = 0.4, shift={(0.4265, -0.1938)}] (-0.605, 0.1128) -- (0.91, 0.6178);   

	\draw [line width = 0.8, color = red] (1.0, 0.67) +(-1.5, -0.5) arc (310: 645: 0.3) node[pos = 0.3, shift = {(-0.3, 0.7)}, black] {};   
	\draw [line width = 0.8, color = red] (1.044, 0.937) +(-1.5, -0.5) arc (130: 465: 0.3) node[pos = 0.6, shift = {(0.35, -0.6)}, black] {};   
	\node[right] at (1.9, 0.55) {$=$};
\end{scope}

\begin{scope}[shift={(3.3, 0.1)}]
    \draw [fill = white!82!blue]  (0.25, 0.5) -- (1.85, -0.5) -- (1.85, 1) -- (0.25, 2) -- cycle;   
    \draw [line width = 0.8, color = red] (1.0, 0.67) arc (310: 645: 0.3) node[pos = 0.5, shift = {(-0.3, 0.35)}, black] {};   
    \draw [line width = 0.8, color = red] (1.044, 0.937) arc (130: 465: 0.3) node[pos = 0.5, shift = {(0.35, -0.3)}, black] {};   
\end{scope}

\end{tikzpicture}
\caption{\label{fig: cylinder_0}%
Intersection/braiding of two-dimensional endable operators in the 4d bulk, as detected by equivalent configurations. From left to right: intersecting disks (the intersection is marked by a yellow dot); braiding semi-infinite cylinders; cylinders terminating on the topological boundary along braiding lines; braiding boundary line operators. This reproduces the braiding between symmetry defects (red circles) of the QFT$_3$ of interest.}
\end{figure}
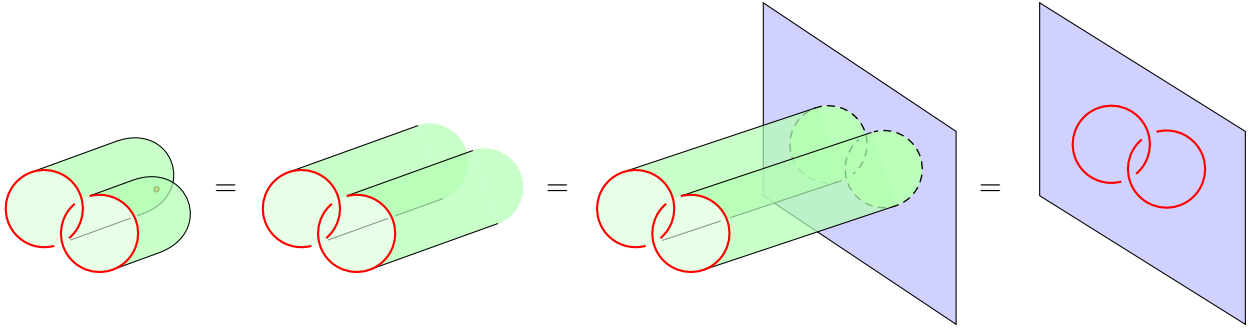

With the goal of clarifying these issues, in this work we study the case of 3d Yang--Mills--Chern--Simons theories in detail. In Section~\ref{section: symTFT in QFT}, focusing for definiteness on the $\su (N)$ gauge algebra, we conduct a field-theoretical analysis and confirm that, in fact, the gauged version of the anomaly inflow SPT phase {\it is} the 4d Symmetry TFT for those theories. Even when the latter is still an SPT, we notice that there can exist inequivalent sets of topological edge modes one can couple it to at the topological boundary, hence re-establishing the one-to-one correspondence with global variants of the 3d theories. It turns out that one can resolve both puzzles by broadening the class of bulk topological operators under scrutiny (similarly to the $G$-crossed braided tensor categories of \cite{Barkeshli:2014cna}) to include ``endable'' operators, \ie{} extended operators in which one can cut a hole. In the language of higher categories, there exist (possibly higher) morphisms between such operators and the identity operator. Exploiting configurations in which the operators are placed along non-intersecting semi-infinite cylinders (as in the middle left portion of Fig.~\ref{fig: cylinder_0}) one can provide a good definition of braiding between the boundaries of the holes. By smoothly closing the open semi-infinite ends of the cylinders (Fig.~\ref{fig: cylinder_0} left) one realizes that such a braiding is equivalent to a contact interaction between the endable operators, for instance placed on intersecting disks. 

By including the behavior of such endable operators in the definition of topological boundary conditions, one recovers the full classification that encompasses the inequivalent edge modes. This can be formalized in terms of a generalized concept of {\it Lagrangian algebra} that includes the endable operators. One could think of the semi-infinite cylinders as terminating on a topological boundary (as in the middle right portion of Fig.~\ref{fig: cylinder_0}). In this way, the braiding between holes is seen to capture the braiding between the edges of the cylinders on the boundary (Fig.~\ref{fig: cylinder_0} right). Such edges are precisely the symmetry defects of the QFT$_3$ that braid among themselves, thus solving the first puzzle as well.%
\footnote{We expect our construction to yield a 4d TQFT that is equivalent to the Crane--Yetter one \cite{Crane:1993if}. Besides, YM-CS theories enjoy infrared dualities originating from the exact level/rank dualities of the corresponding pure Chern--Simons theories in the IR. For instance, $SU(N)_k$ is dual to $U(k)_{-N}$ so that both have a $\bZ_N$ 1-form symmetry with the same anomaly. It follows that they also enjoy the same Symmetry TFT. This generalizes to other groups.}

In the second part of the paper we leverage holography in type IIB string theory to verify our claims. At first glance this should not be expected since holography is a different framework, in which a boundary QFT$_d$ is encoded in a $(d+1)$-dimensional bulk and the bulk theory has local dynamics (and it also includes gravity). It is possible, however, to single out a topological sector of the bulk theory that describes the symmetry properties of the QFT$_d$. Due to the bulk background metric, at large $r$, \ie{} in the near-boundary region, the bulk field kinetic terms are suppressed and the bulk action is dominated by topological terms \cite{Witten:1998wy}. This is the region one should focus on to distill all RG-invariant properties of the dual gauge theory,  which are described by the Symmetry TFT. From a  holographic point of view, the $(d+1)$-dimensional slab in Fig.~\ref{fig: slab} can then be thought of as describing a tiny portion of the dual bulk, parametrically close to the holographic boundary \cite{Heckman:2024oot}.  All dynamical data --- the RG flow and the IR dynamics of the QFT --- are instead encoded in the physical boundary.

The model we consider is based on a system of D3/D7-branes in type IIB string theory compactified on a supersymmetry-breaking circle, proposed long ago in \cite{Witten:1998zw, Fujita:2009kw} as a holographic model of (the universality class of) $SU(N)_k$ YM-CS theory at large $N$.
In Section~\ref{section: holographic symTFT} we extract its topological sector%
\footnote{The analysis of global variants of the ABJM supersymmetric Chern--Simons--matter theories and of the corresponding boundary conditions for a topological sector of the holographic dual in type IIA string theory was performed in \cite{Bergman:2020ifi}.}
  and verify that it agrees with the Symmetry TFT proposed in Section~\ref{section: symTFT in QFT}. We also provide a description of symmetry defects and charged operators associated to the 1-form symmetry of $SU(N)_k$ YM-CS theory in terms of string states (D-branes, fundamental strings, and bound states thereof), and show that they behave as expected in terms of their (non-)topological nature and braiding properties.
Finally, Section~\ref{section: holographic RG} contains a discussion on how holography captures the way line operators of the UV YM-CS theory evolve along the RG flow from being a set of topological~/~non-topological operators to becoming all topological in the deep IR, where the full non-invertible 1-form symmetry of pure CS theory is expected to be realized. 
In addition to providing an independent check of our field theory analysis, these results provide further evidence that the holographic model of \cite{Witten:1998zw, Fujita:2009kw} is indeed in the same universality class as three-dimensional YM-CS theory (see \eg{} \cite{Rey:2008zz, Hong:2010sb, Argurio:2020her, Argurio:2022vtz} for further results supporting this claim).


\section{Symmetry TFT of \tps{\matht{\mathfrak{su}(N)_k}}{su(N)k} Yang--Mills--Chern--Simons theory}
\label{section: symTFT in QFT}

Three-dimensional Yang--Mills--Chern--Simons (YM-CS) theory with gauge algebra $\su(N)$ and level $k$ admits multiple global variants (or global forms), all related by ``topological manipulations'', \ie{} by gauging some finite internal symmetry of the theory. As we recall in Appendix~\ref{app: su(N)k Chern-Simons}, for given $N$ and $k$ the global variants have gauge group $\bigl( SU(N) / \bZ_p \bigr){}_k$ where $p$ is a divisor of $N$ that satisfies an extra condition. For bosonic theories, namely theories that do not require the existence of a spin structure on the spacetime manifold, $k$ must be a multiple of
\be
\label{condition on k bosonic}
k_0^\text{bos} = \begin{cases} \dfrac{p}{\gcd\bigl( p, \frac Np \bigr)} &\text{for $p$ odd} \,, \\[1.2em]
\dfrac{2p}{\gcd\bigl( 2p, \frac Np \bigr)} & \text{for $p$ even} \,. \end{cases}
\ee
For spin theories, namely theories that require and depend on a spin structure on spacetime, $k$ must be a multiple of
\be
\label{condition on k spin}
k_0^\text{spin} = \frac{p}{\gcd \bigl( p , \frac Np \bigr)} \,.
\ee
This result can be obtained in (at least) two ways. One is to study for which groups the level $k$ Chern--Simons term is well defined (we review it in Appendix \ref{app: level quant and variants}). The other way is to start from the theory with simply-connected gauge group $SU(N)_k$. This theory has a $\bZ_N$ 1-form symmetry, implemented by topological Gukov--Witten (GW) operators $U_m[\gamma]$ supported on one-dimensional lines $\gamma$ and labeled by elements $g = \exp\bigl( \frac{2\pi i}N m \bigr)$, with $m \in \bZ_N$, in the center of the gauge group. Among charged operators there are the \emph{non-topological} Wilson lines defined as
\be
W_{\mathcal{R}} = \Tr_{\mathcal{R}} \Pexp \biggl( i \oint A \biggr) \,,
\ee
where $\mathcal{R}$ is a representation of $SU(N)$, $A$ is the $SU(N)$ gauge field, and $\Pexp$ is the path-ordered exponential. Such lines have charge $q_{\mathcal{R}}$ equal to the $N$-ality (number of boxes in the Young diagram) of the representation $\mathcal{R}$.%
\footnote{In the topological pure CS theory, the Gukov--Witten operators $U_m[\gamma]$ coincide with the Abelian Wilson lines whose defining representation is labeled by a rectangular Young diagram with $k$ columns and $m$ rows. In the full YM-CS theory, instead, all Wilson lines are non-topological and are distinct from the Gukov--Witten operators.}
The 1-form symmetry has an 't~Hooft anomaly captured by the mutual braiding $\bB_{mm'}$ between GW operators and by the spin $\theta_m$, both taking values in $U(1)$:
\be
\label{eq: 3d brainding and spin}
\bB_{mm'} = \exp\biggl( \frac{2\pi i k}N \, mm' \biggr) \,,\qquad\qquad \theta_m = \exp\biggl( \frac{\pi i k}N\, m \, (N-m) \biggr) \,.
\ee%
\begin{figure}[t]
\centering
\begin{tikzpicture}
	\draw [thick] (0.14, 0.5) node[above left] {\small$U_m[\gamma]$} arc [start angle = 180, end angle = 157, radius = 5];
	\filldraw [white] (0.21, 1.34) circle [radius = 0.12];
	\draw [thick, black, rotate around = {-10: (0.23, 2.12)}] (0.23, 2.12) arc [start angle = 110, end angle = 451, x radius = 1.1, y radius = 0.4] node[pos = 0.65, below, shift={(0,-0.1)}] {\small$U_{m'}[\gamma']$};
       \node[right] at (2.3, 1.6) {\large$ = \quad \bB_{mm'} $};
\begin{scope}[shift={(4.9,0)}]
	\draw [thick] (0.14, 0.5) node[above left] {\small$U_m[\gamma]$} arc [start angle = 180, end angle = 157, radius = 5];
	\draw [thick, black, rotate around = {-10: (1.4, 1.95)}] (1.4, 1.95) arc [start angle = 110, end angle = 470, x radius = 1.1, y radius = 0.4] node[pos = 0.65, below, shift={(0,-0.1)}] {\small$U_{m'}[\gamma']$};
\end{scope}
\begin{scope}[shift={(11.5,0)}]
	\draw [thick] (0, 0.5) node[above left] {\small$U_m[\gamma]$} -- (0, 1.4);
	\draw [thick] (0, 1.4) arc [start angle = 180, end angle = -135, radius = 0.25];
	\draw [thick] (0, 1.65) -- (0, 2.3);
	\node [right] at (0.9, 1.4) {\large$ = \;\; \theta_m$};
	\draw [thick] (2.5, 0.5) node[above right] {\small$U_m[\gamma]$} -- (2.5, 2.3);
\end{scope}
\end{tikzpicture}
\caption{\label{fig: braiding and spin}%
Topological manipulations of GW operators that define the braiding $\bB_{mm'}$ and the spin $\theta_m$.}
\end{figure}%
They are determined by the manipulations in Fig.~\ref{fig: braiding and spin}, and are related by \cite{Kitaev:2005hzj}
\be
\bB_{mm'} = \frac{\theta_m \, \theta_{m'} }{ \theta_{m + m'}} \,.
\ee
One way to think of this anomaly is that the topological operators $U_m[\gamma]$ are both symmetry defects and charged objects under the symmetry. As a consequence, they are not gauge invariant under background gauge transformations when a background gauge field $\cB \in H^2(X_3;\bZ_N) $ is turned on (here $X_3$ is the spacetime manifold). In the bosonic case, writing the spin as $\theta_m = \exp\bigl( -\pi i \ell \, \frac{m^2}{N} \bigr)$ \cite{Hsin:2018vcg}, the anomaly is quantified by an integer $\ell \in \bZ_{2N}$ that is given by
\be
\ell = \begin{cases} k+N &\text{for $k$ odd} \,, \\ k &\text{for $k$ even} \,. \end{cases}
\ee
Notice that $N \ell$ is always even. In the context of spin theories, instead, one identifies $\ell \sim \ell + N$ \cite{Hsin:2018vcg}, therefore if we regard $SU(N)_k$ as a spin theory then $\ell = k$ in $\bZ_N$.
The anomaly inflow action for the 1-form symmetry is
\be
\label{eq: inflow}
\cA_\text{inflow}[\cB] = \frac{2\pi i \ell }{2N} \int_{\cM_4} \! \mathcal{P}(\cB) \,,
\ee
where $\cM_4 $ is a four-dimensional bulk manifold with $\partial \cM_4 = X_3$, whilst $\mathcal{P}$ is the  Pontryagin square operation. Because of the anomaly, the $\bZ_N$ 1-form symmetry cannot be gauged in general. However, there can exist a $\bZ_p$ subgroup (generated by a subset of GW lines) which is anomaly free and, upon proper gauging, leads to another global variant \cite{Moore:1989yh}. This anomaly-free subgroup exists precisely when the conditions \eqref{condition on k bosonic} and \eqref{condition on k spin} are satisfied. Note that the anomaly cancellation condition is different in the bosonic and spin cases \cite{Gaiotto:2014kfa}. See Appendix~\ref{app: level quant and variants} for details.

Our goal is to characterize the four-dimensional Symmetry TFT which describes all global variants and topological manipulations of YM-CS theory with gauge algebra $\su(N)$ and level $k$. If, as in our case, one starts from a theory with an invertible symmetry, a constructive way to produce the Symmetry TFT is to couple the $d$-dimensional boundary theory to a background gauge field $\cB$ with $(d+1)$-dimensional anomaly inflow action and make $\cB$ dynamical. Placing the system on a $(d+1)$-dimensional slab and imposing (topological) Dirichlet boundary conditions for $\cB$ on one boundary, one recovers the original theory on the other boundary. Consequently, the Symmetry TFT emerges as the gauged version of the anomaly inflow action $\cA_\text{inflow}[\cB]$. In our case, one obtains a $\bZ_N$ Dijkgraaf--Witten theory \cite{Dijkgraaf:1989pz} with twist $\ell$, which admits a description in terms of $U(1)$ one-form and two-form gauge fields $C$ and $B$, respectively, with Euclidean action \cite{Maldacena:2001ss, Banks:2010zn, Kapustin:2014gua}
\be
\label{eq: symTFT}
S_\text{SymTFT} = \int_{\cM_4} \biggl[ \frac{iN}{2\pi} \, B \wedge dC + \frac{iN\ell}{4\pi} \, B\wedge B \biggr] \,.
\ee
This theory has been studied in \cite{Kapustin:2014gua} (see also \cite{Putrov:2016qdo, Ye:2014oua, Zhang:2023ynd}). Let us review some properties of the theory in (\ref{eq: symTFT}). The gauge transformations are
\be
\label{eq: gauge transf}
B \,\rightarrow\, B + d\lambda_B \,,\qquad\qquad C \,\rightarrow\, C + d\lambda_C - \ell \, \lambda_B \,,
\ee
under which the action transforms by a local total derivative
\be
\delta S_\text{SymTFT} = \int_{\cM_4} \biggl[ \frac{iN}{2\pi} \, d\lambda_B \wedge dC - \frac{iN\ell}{4\pi} \, d\lambda_B \wedge d\lambda_B \biggr] \,.
\ee
Recalling that gauge fields and transformations satisfy Dirac quantization conditions, this variation is an integer multiple of $2\pi$ on generic (orientable) closed manifolds if $N\ell \in 2\bZ$, and on closed spin manifold for any $N\ell \in \bZ$. Since these conditions are satisfied, the theory is gauge invariant.

The gauge transformations and the equations of motion imply that the Wilson surface of $B$ on a closed surface $\Sigma_2$,
\be
\label{def of W_B}
W_B[\Sigma_2] = \exp\biggl( i \!\int_{\Sigma_2} \! B \biggr) \,,
\ee
is gauge invariant and topological. On the contrary, the Wilson line of $C$, $W_C[\gamma] = \exp{(i \!\int_\gamma C)}$, is not and needs to stay at the boundary of $(W_B)^\ell$ in order to be well defined. In  other words 
\be
\label{nongenl}
\cW_C[D_2] \equiv W_C[\gamma] \; (W_B)^\ell[D_2] = \exp \biggl[ i \! \int_{D_2} \bigl( dC + \ell \, B \big) \biggr] \,,
\ee
where $D_2$ is an open surface with $\partial D_2 = \gamma$, is the correct gauge-invariant topological object. Operators like $\cW_C[D_2]$ are sometimes called non-genuine line operators because, as a matter of fact, they explicitly depend on the surface $D_2$ attached to the line $\gamma$. The equations of motion obtained from the sum over gauge fluxes in (\ref{eq: symTFT}) impose that 
\be
(W_B)^N[\Sigma_2] = 1 \,, \qquad\qquad \Bigl( W_C[\gamma] \; (W_B)^\ell[D_2] \Bigr)^N = 1 \,.
\ee
They imply that $(W_C)^{N/g}[\gamma]$, where
\be
\label{def of g}
g = \gcd(N,\ell) = \gcd(N,k) \;,
\ee
is a genuine line operator which does not require an open surface $D_2$. On the other hand, in the language of Section~\ref{sec: intro}, the surface operator $(W_B)^g$ and its powers are \emph{endable} because one can cut a topological hole in them by terminating them on suitable powers of $W_C$. Endable topological operators are often regarded as trivial, because one can cut holes in them and shrink them to a point. As a consequence, in our case the standard lore would be to consider as nontrivial bulk operators only the powers $(W_B)^n$ and $(W_C)^{n \, N/g}$ for $n<g$. However, endable operators can have contact interactions with other operators (namely, their correlation functions can be nontrivial if the operators cross each other), thus they are not completely trivial. This will be important in the following.


\subsection{Topological boundary conditions via Lagrangian algebras}
\label{subsection: symTFT TBC}

We expect a one-to-one correspondence between topological boundary conditions of the Symmetry TFT and global variants of the boundary theory. Let us then analyze such boundary conditions. For Abelian TQFTs, the topological boundary conditions are in correspondence with \emph{Lagrangian algebras}, namely with maximal subsets of the topological operators that are closed under fusion, have trivial braiding with each other, and in the bosonic case have trivial spin, $\theta=1$.%
\footnote{In the spin case, the triviality of the spin means $\theta=\pm 1$, which is already guaranteed by the triviality of the braiding. In the bosonic case, we need to further impose $\theta = +1$.}
However, one has to be careful because when the TQFT includes some SPT phases there are operators that couple via contact interactions rather than by braiding. For instance, when $g=1$ the TQFT in (\ref{eq: symTFT}) is an SPT phase: all its operators are either non-genuine or endable, meaning that the Hilbert space is one-dimensional. Therefore, it seems that there is no nontrivial braiding to define. Indeed the theory has a unique state on any spatial manifold (because, by the state/operator correspondence for TQFTs, only genuine non-endable operators create nontrivial states), and identifying boundary conditions with states one would conclude that there exists a unique boundary condition. Instead, it turns out that an SPT phase can admit multiple inequivalent boundary conditions, corresponding to the global variants of the boundary theory.

One way to understand this fact is to notice that if the Symmetry TFT is described by an SPT phase, then the operators which implement the symmetry must be endable. They also produce the nontrivial phases of the partition function when a network of them is inserted. Correlation functions in this theory are all trivial, but for contact terms. In the presence of a boundary, one can broaden the set of operators by including the open surfaces that terminate on the boundary, as in the middle right of Fig.~\ref{fig: cylinder_0}. Alternatively, by pushing the topological boundary to infinity, one can think of them as non-intersecting semi-infinite cylinders, as in the middle left of Fig.~\ref{fig: cylinder_0}. In both presentations, such endable operators can now link between each other \emph{without} intersecting, and hence define a braiding.%
\footnote{Usually, braiding in a $d$-dimensional manifold is defined through the linking between two closed $p$- and $q$- surfaces such that $p+q=d-1$. In this case, we are considering two 2-surfaces in 4d. However, the presence of a boundary allows for an extended definition of braiding between the non-intersecting endable operators: in any slice parallel to the boundary, they braid according to the usual definition.}
This reproduces the usual braiding between the boundary lines, as in the right of Fig.~\ref{fig: cylinder_0}.  
Notice that, instead of letting the endable surfaces run to infinity, one can smoothly close them: then, the linking number of their boundaries equals their intersection number, as in the left of Fig.~\ref{fig: cylinder_0}. In this way, we see that the contact interaction between endable surfaces, which captures the SPT phase, is responsible for the braiding of the corresponding boundary lines.
One is thus led to specify boundary conditions in terms of a generalized definition of Lagrangian algebra, as a maximal set of \emph{both} genuine and non-genuine topological operators which do not braid with each other (plus a condition on the spin in the bosonic case).

Let us apply this definition to the TQFT in (\ref{eq: symTFT}). The full set of topological operators is given by
\be
(W_B)^m[\Sigma_2] \,,\qquad (\cW_C)^r[D_2] \,,\qquad \text{with } m,r \in \bZ_N \,.
\ee
They satisfy the following braiding relations
\be
\label{braiding in SymTFT}
\bB\Bigl( (W_B)^m, (\cW_C)^r \Bigr) = \exp\biggl( \frac{2\pi i}N \, m r \biggr) \,,\qquad \bB\Bigl( (\cW_C)^{r_1}, (\cW_C)^{r_2} \Bigr) = \exp\biggl( \frac{2\pi i \ell}N \, r_1 r_2 \biggr) \,.
\ee
Notice that correctly, since an endable surface and a genuine line do not braid, the first braiding is trivial when $m$ is a multiple of $g$ and $r$ is a multiple of $N/g$, and the second braiding is trivial when either $r_1$ or $r_2$ is a multiple of $N/g$. We can also associate a spin to the operators $(\cW_C)^r$ since their bulk has codimension 2:
\be
\label{spin in SymTFT}
\theta_r = \exp\biggl( - \frac{\pi i \ell}N \, r^2 \biggr) \,.
\ee
Notice that the genuine line operators $(W_C)^{a N/g}[\gamma]$ have spin $\theta_{aN/g} = \pm1$, consistently with the fact that they have codimension 3. We construct a Lagrangian algebra with the subset of operators $\{(W_B)^{m_i},(\cW_C)^{r_i}\}$. The integers $\{m_i, r_i\}$ take values in $\bZ_N$. Triviality of the first braiding in (\ref{braiding in SymTFT}) and maximality impose that $m_i \in p\,\bZ$, $r_i \in q\,\bZ$ for some integers $p,q$ such that $N = pq$. Triviality of the second braiding in (\ref{braiding in SymTFT}) imposes that $q^2\ell$ be a multiple of $N$, namely
\be
\label{condition on ell spin}
\ell \; \text{ must be a multiple of } \; \frac{p}{\gcd\bigl( p, \frac Np \bigr)} \,.
\ee
In the bosonic case, triviality of the spin in (\ref{spin in SymTFT}) imposes the stronger constraint that $q^2\ell$ be a multiple of $2N$, namely
\be
\label{condition on ell bosonic}
\ell \; \text{ must be a multiple of } \; \frac{2p}{\gcd\bigl( 2p, \frac Np \bigr)} \,.
\ee
In the spin case $\ell \in \bZ_N$ and $\ell = k \text{ mod } N$. We immediately see that the condition on $\ell$ in (\ref{condition on ell spin}) is equivalent to the condition on $k$ in (\ref{condition on k spin}).
Also in the bosonic case, where $\ell \in \bZ_{2N}$, one can verify with a little bit of algebra that the condition on $\ell$ in (\ref{condition on ell bosonic}) is equivalent to the condition on $k$ in (\ref{condition on k bosonic}).
Thus, the set of Lagrangian algebras is in one-to-one correspondence with the global variants of the boundary theory. Indeed the topological surface operators $(W_B)^m[\Sigma_2]$ modulo the Lagrangian algebra, when laid on the boundary, become the symmetry defects of a $\bZ_p$ 0-form symmetry of the boundary theory that we identify with the magnetic symmetry of $SU(N)/\bZ_p$. We can thus identify the integer $p$ introduced here with the one introduced before. In the following we will sometime use the notation $\bZ^{[j]}_\#$ to indicate a $j$-form symmetry.

Not all of the $\bZ_p$ 0-form symmetry acts faithfully. In principle a monopole operator with unit charge is the endpoint of $(\cW_C)^{q}$. However, the latter is in general not a genuine line. As a consequence, the monopole operator is not genuine itself, \ie{} it is not a gauge-invariant operator of the theory on the physical boundary. The genuine lines that can end on the topological boundary are $(\cW_C)^{rq}$ with $r \in \frac{p}{\gcd(p,\ell)} \, \bZ$.%
\footnote{The minimal charge of a genuine line that can end on the boundary is $\operatorname{lcm}(N/g, q) = N / \gcd(p,g) = pq / \gcd(p,\ell)$.}
It follows that only the monopoles of charge multiple of $\frac{p}{\gcd(p,\ell)}$ are gauge invariant, and thus the subgroup $\bZ_{p/\gcd(p,\ell)} \subset \bZ_p$ of the 0-form symmetry does not act on anything.

Similarly, we can determine the electric 1-form symmetry of a global variant $\bigl( SU(N)/\bZ_p \bigr){}_k$. Its symmetry defects should be the topological operators $(\cW_C)^r[D_2]$ laid on the boundary, modulo the Lagrangian algebra. This operation would seem to give a $\bZ_q$ 1-form symmetry for the boundary theory, generated by the operators with $r \in \bZ_q$. However, these operators also depend on the surfaces of $B$ they are attached to, which appear with the integer power $r \ell$. In order for $(\cW_C)^r$ to be a genuine boundary line, we have to trivialize its surface dependence by requiring that $(W_B)^{r \ell}$ is in the Lagrangian algebra. This imposes $r\ell$ to be a multiple of $p$, namely $r = \frac{p}{\gcd(p,\ell)} \, s$ with $s \in \bZ_L$, where%
\footnote{The consistency conditions (\ref{condition on ell spin}) or (\ref{condition on ell bosonic}) guarantee that $q$ is a multiple of $p/\gcd(p,\ell)$, see Appendix~\ref{app: SU(N)_k/Z_p 1-form sym}.}
\be
\label{L from lagalg}
L = \frac{q}{p/\gcd(p,\ell)} = \frac{N\gcd(p,k)}{p^2} \,,
\ee
where we used the fact that $\gcd(p,\ell) = \gcd(p,k)$.
The defects $(\cW_C)^{\frac{p}{\gcd(p,\ell)} \, s}$ laid on the boundary are the genuine lines realizing the $\bZ_L$ electric 1-form symmetry of the global variant $\bigl( SU(N)/\bZ_p \bigr){}_k$. Obviously, the whole $\mathbb{Z}_L$ symmetry acts faithfully. The braiding between the generating lines can be determined from \eqref{braiding in SymTFT} to be
\be
\label{braiding from lagalg}
\bB\Bigl( (\cW_C)^{\frac{p}{\gcd(p,\ell)}s}, (\cW_C)^{\frac{p}{\gcd(p,\ell)}s'} \Bigr) = \exp\biggl[ 2\pi i \, \frac{\ell}{L \gcd(p, \ell) } \, s s' \biggr] = \exp\biggl[ 2\pi i \, \frac{p^2 k}{N \gcd(p, k)^2 } \, s s' \biggr]\,.
\ee
These results precisely agree with the order of the symmetry and the braiding between topological Abelian lines in $\bigl( SU(N)/\bZ_p \bigr){}_k$ CS theory (see Appendix~\ref{app: SU(N)_k/Z_p 1-form sym}), and in the special case where $p=1$ they reproduce (\ref{eq: 3d brainding and spin}).


\subsection{Topological boundary conditions via edge modes}
\label{subsection: symTFT edgemodes}

An alternative way to describe the inequivalent topological boundary conditions is in terms of topological edge modes. When the theory in \eqref{eq: symTFT} is placed on a manifold $\cM_4$ with boundary, under the gauge transformations (\ref{eq: gauge transf}) the action transforms by a boundary term:
\be
\label{eq: bdry gauge var}
S_\text{SymTFT} \;\rightarrow\; S_\text{SymTFT} + \int_{\partial \cM_4} \biggl[ \frac{iN}{2\pi} \, \lambda_B \wedge dC - \frac{iN\ell}{4\pi} \, \lambda_B \wedge d \lambda_B \biggr] \,.
\ee
In order to make the theory gauge invariant, we need to impose boundary conditions which trivialize such a variation. This can be achieved by either constraining the gauge fields and their variations on the boundary, or introducing dynamical edge modes that couple to the bulk fields and enforce boundary conditions. While the former method is conceptually more straightforward, the latter makes it clearer what is the full list of boundary conditions and elucidates the fate of bulk operators pushed to the boundary. 

We introduce a boundary one-form degree of freedom $\phi$ (a standard $U(1)$ gauge field) with topological Chern--Simons action, coupled to the bulk \eqref{eq: symTFT} by the following action:
\be
\label{action 3d4d with edge modes}
S_\text{3d/4d} = S_\text{SymTFT} + \int_{\partial \cM_4} \biggl[ \frac{iq}{2\pi} \, \phi \wedge dC - \frac{iy}{4\pi} \, \phi \wedge d\phi \biggr] \,,
\ee
and with gauge transformation
\be
\phi \,\rightarrow\, \phi + d\lambda_\phi - p \, \lambda_B \,,
\ee
which supplements (\ref{eq: gauge transf}). Here $p$, $q$, $y$ are arbitrary integer parameters, although $y$ is constrained to be even for a bosonic edge mode. Taking into account that $\partial\cM_4$ has no boundary, the coupled 3d/4d system is gauge invariant if and only if
\be
pq = N \qquad\text{and}\qquad yp = q\ell \,.
\ee
In the spin case, this requires that $p$ be a divisor of $N$ and $q\ell$ a multiple of $p$, which is equivalent to (\ref{condition on ell spin}) and in turn to (\ref{condition on k spin}). In the bosonic case where $y$ has to be even, the requirement is that $p$ be a divisor of $N$ and $q\ell$ a multiple of $2p$, which is equivalent to (\ref{condition on ell bosonic}) and in turn to (\ref{condition on k bosonic}). So we see that in both cases the set of topological edge modes is in one-to-one correspondence with the global variants of the physical boundary theory. Notice that $p=1$ is always a solution for any $N$ and $k$ (or $\ell$), because the global variant $SU(N)_k$ always exists both as a bosonic and a spin theory. It is worth stressing that even when $\gcd(N,\ell) =1$ so that the Symmetry TFT is a four-dimensional SPT phase, there can be inequivalent boundary conditions corresponding to different global variants of the physical boundary theory.%
\footnote{For instance, consider $\su(N)_k$ with $N=16$ and $k$ odd. In the spin case, $\ell = k \text{ mod }N$ therefore $\ell$ is odd and $\gcd(N,\ell)=1$. The spin global variants $\bigl( SU(N)/\bZ_p \bigr){}_k$ exist for $p = 1,2,4$. In the bosonic case, $\ell = k + N \text{ mod } 2N$ therefore $\ell$ is again odd and $\gcd(N, \ell)=1$. The bosonic global variants exist for $p = 1,2$.}

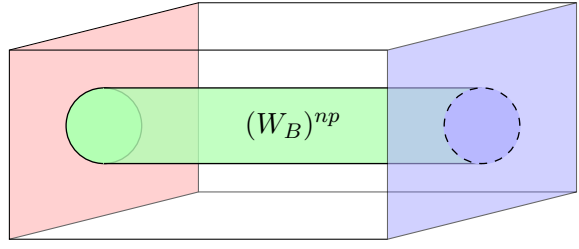
\begin{figure}[t]
\centering
\begin{tikzpicture}[scale=2.5]
	\draw (1, 0.25) -- (3, 0.25); \draw (1, 1.25) -- (3, 1.25);   
	\filldraw [fill = red!30!white] (0,0) -- (1,0.25) -- (1,1.25) -- (0,1) -- cycle;   
	\fill [white, opacity = 0.4] (0.01, 1) -- (2, 1) -- (2.98, 1.247) -- (1, 1.247) -- cycle; 
	\fill [white, opacity = 0.4] (0.0028 ,0) -- (2,0) -- (2,1) -- (0.0028 ,1) -- cycle; 
	\draw (0,0) -- (2,0); \draw (0,1) -- (2,1);   
	\draw [line width = 0.5, fill = white!70!green] (0.5, 0.6) circle [radius = 0.2];  
	\fill [white!70!green, opacity=0.8] (0.5, 0.4) -- (2.5, 0.4) -- (2.5, 0.8) -- (0.5, 0.8) -- cycle;  
	\draw [line width = 0.5, color=black] (0.5, 0.4)--(2.5,0.4);
	\draw [line width = 0.5, color=black] (0.5, 0.8)--(2.5,0.8);
	\node at (1.5, 0.6) {$(W_B)^{np}$};
	\draw [fill = white!70!blue, opacity=0.6] (2,0) -- (3,0.25) -- (3,1.25) -- (2,1) -- cycle;  
	\filldraw [white!72!blue] (2.5, 0.6) circle [radius = 0.2007];   
	\draw [line width = 0.5, dashed] (2.5, 0.6) circle [radius = 0.2];
\end{tikzpicture}
\caption{\label{fig: slab configuration}%
Slab description of a Wilson line of 3d $\bigl( SU(N)/\bZ_p \bigr){}_k$ YM-CS theory (with $N$-ality $np$), as an operator $(W_B)^{np}$ stretched between the physical boundary on the left (where the QFT$_3$ lives) and the topological boundary on the right.}
\end{figure}

The equations of motion give the following boundary conditions:
\be
\label{boundary conditions with edge modes}
p\, B \big|_{\partial\cM_4} = - d\phi \,,\qquad\qquad q\, dC \big|_{\partial\cM_4} = y\, d\phi \,.
\ee
The first one implies that a multiple of $B$ is pure gauge on the boundary and is thus a (partial) Dirichlet boundary condition. In the special case that $p=1$ corresponding to $SU(N)_k$ YM-CS theory, those are Dirichlet boundary conditions for $B$ and free (Neumann) for $C$. As already noticed above using the language of Lagrangian algebras, the bulk operators
\be
(W_B)^m[\Sigma_2] \qquad\text{with $m \in \bZ_p$}
\ee
remain nontrivial when laid on the 3d topological boundary and become the symmetry defects of a $\bZ_p$ 0-form symmetry. On the contrary, the operators $(W_B)^{np}$ get trivialized at the boundary because of (\ref{boundary conditions with edge modes}), but they can consistently be opened to terminate on the Wilson lines of $\phi$, since the operators
\be
\exp\biggl[ in \biggl( \, \int_{D_2} \! p \, B + \int_{\partial D_2} \phi \biggr)\biggr] \qquad\text{with $n \in \bZ$}
\ee
are gauge invariant. Here $D_2$ is an open surface in the bulk, whose boundary $\partial D_2 \subset \partial \cM_4$. In the picture of the Symmetry TFT on a slab one has the usual configurations of operators stretched between the physical and the topological boundary, that represent non-topological Wilson lines in the 3d physical YM-CS theory, see Fig.~\ref{fig: slab configuration}.

On the other hand, one can define genuine line operators stuck at the topological boundary along a curve $\hat\gamma\subset \partial \cM_4$, of the form
\be
\label{boundary line operator}
\bigl( \wh W_C \bigr)^s[\hat\gamma] = \exp\biggl[ i \,\frac{s}{L} \int_{\hat\gamma} \Bigl( q \, C - y \, \phi \Bigr) \biggr] \qquad\qquad\text{with $s \in \bZ_L$} \,,
\ee
where
\be
\label{def of L}
L = \gcd(q,y) = \frac{q}{p/\gcd(p,\ell)} = \frac{N \gcd(p,k) }{ p^2} \,.
\ee
The order $L$ of the group follows from the boundary equations of motion, \ie{} from the sum over boundary fluxes of $\phi$ which sets $\int_\partial \bigl( qC - y\phi \bigr) \in 2\pi\bZ$. In expressing it, we used the relations $yp=q\ell$, $N=pq$, and the fact that $\gcd(p,\ell) = \gcd(p,k)$. These boundary line operators are the symmetry defects of a $\bZ_L$ electric 1-form symmetry, reproducing the previous result in \eqref{L from lagalg}.

The equations of motion also determine the braiding between boundary line operators as
\be
\label{boundarybraiding}
\bB\Bigl( \bigl( \wh W_C \bigr)^s \,,\, \bigl( \wh W_C \bigr)^{s'} \Bigr) = \exp\biggl[ 2\pi i \, \frac{\ell}{L \gcd(p, \ell) } \, s s' \biggr] = \exp\biggl[ 2\pi i \, \frac{p^2 k}{N \gcd(p, k)^2 } \, s s' \biggr]\,,
\ee
which, again, fully agrees with our findings in \eqref{braiding from lagalg}.

Unlike in more standard instances of Symmetry TFTs, most of the boundary lines do not arise from bulk lines pushed to the boundary. Indeed, the bulk TQFT has $g=\gcd(N,\ell)$ topological line operators generated by $(W_C)^{N/g}$. These bulk lines cannot braid among themselves for dimensional reasons, and they continue not to braid even when pushed to the boundary. In fact, at the boundary they can be re-expressed as%
\footnote{From the previous discussion, the exponent on the right-hand side is $[N/g]\bigl[ \gcd(p,\ell)/ p \bigr]$. This can be rewritten as in (\ref{WCrelationbulkboundary}) using two identities: $g = \gcd(p,\ell) \gcd\bigl( q,\, \ell / \gcd(p, \ell) \bigr)$, and $\gcd \bigl( q, \, \ell/\gcd(p,\ell) \bigr) = \gcd \bigl( L,\, \ell / \gcd(p,\ell) \bigr)$ which follows from writing $q = L \cdot p/\gcd(p,\ell)$.}
\be
\label{WCrelationbulkboundary}
(W_C)^{N/g} = (\wh W_C)^{\tfrac{p}{\gcd(p,\ell)} \, \tfrac{L}{\gcd \left( \rule{0pt}{0.6em} L, \; \ell/\gcd(p,\ell) \right)}} \,.
\ee
Note that the power of $\wh W_C$ is such that the dependence on $\phi$ trivializes due to the sum over boundary fluxes of $C$ which sets $q\!\int_\partial \phi  \in 2\pi\bZ$. The operator in (\ref{WCrelationbulkboundary}) and its powers are in general a subset of the boundary lines. By plugging the adequate power in the expression \eqref{boundarybraiding}, one can check that indeed they have trivial braiding among themselves. 

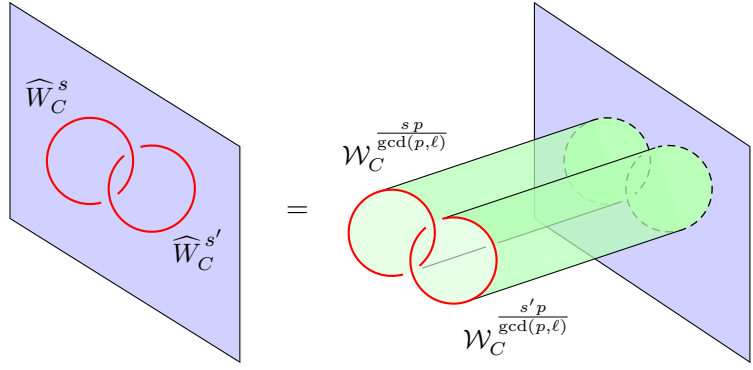
\begin{figure}[t]
\centering
\begin{tikzpicture}[scale=1.9]
	\draw [fill = white!82!blue]  (0.25, 0.5) -- (1.85, -0.5) -- (1.85, 1) -- (0.25, 2) -- cycle;   
	\draw [line width = 0.8, color = red] (1.0, 0.67) arc (310: 645: 0.3) node[pos = 0.5, shift = {(-0.3, 0.35)}, black] {\small$\wh{W}_C^{\raisebox{0.25em}{\scriptsize$\,s$} }$};   
	\draw [line width = 0.8, color = red] (1.044, 0.937) arc (130: 465: 0.3) node[pos = 0.5, shift = {(0.35, -0.3)}, black] {\small$\wh{W}_C^{\raisebox{0.25em}{\scriptsize$\,s'$} }$};   
	\node[right] at (2.1, 0.55) {$=$};

\begin{scope}[shift={(3.6, 0)}]
	\draw [fill = white!82!blue]  (0.3, 0.5) -- (1.8, -0.5) -- (1.8, 1) -- (0.3, 2) -- cycle;   

	\fill [white!73!green] (1.0, 0.67) arc (310: 670: 0.3);   
	\fill [white!73!green] (1.044, 0.937) arc (130: 490: 0.3);   
	\draw [line width = 0.5, densely dashed, dash phase = 2pt] (1.0, 0.67) arc (310: 645: 0.3);   
	\draw [line width = 0.5, densely dashed, dash phase = 1.8pt] (1.044, 0.937) arc (130: 465: 0.3);   

	\fill [white!70!green, opacity = 0.7] (0.715, 1.1855) -- (-0.794, 0.6825) -- (-0.605, 0.1128) -- (0.91, 0.6178) -- cycle;   
	\draw [line width = 0.4, shorten < = 37] (-0.605, 0.1128) -- (0.91, 0.6178);   
	\fill [white!70!green, opacity = 0.7, shift={(0.4265, -0.1938)}] (0.715, 1.1855) -- (-0.794, 0.6825) -- (-0.605, 0.1128) -- (0.91, 0.6178) -- cycle;   

	\fill [white!90!green] (1.0, 0.67) +(-1.5, -0.5) arc (310: 670: 0.3);   
	\fill [white!90!green] (1.044, 0.937) +(-1.5, -0.5) arc (130: 490: 0.3);   

	\draw [line width = 0.4] (0.715, 1.1855) -- (-0.794, 0.6825);   
	\draw [line width = 0.4, shift={(0.4265, -0.1938)}, shorten > = 3] (0.715, 1.1855) -- (-0.794, 0.6825);   
	\draw [line width = 0.4, white!90!green!60!black, shorten < = 7, shorten > = 54] (-0.605, 0.1128) -- (0.91, 0.6178);   
	\draw [line width = 0.4, shift={(0.4265, -0.1938)}] (-0.605, 0.1128) -- (0.91, 0.6178);   

	\draw [line width = 0.8, color = red] (1.0, 0.67) +(-1.5, -0.5) arc (310: 645: 0.3) node[pos = 0.3, shift = {(-0.3, 0.7)}, black] {\small$ \cW_C^{\,\raisebox{0.6em}{\scriptsize$\frac{s \, p}{\gcd(p,\ell)}$} }$};   
	\draw [line width = 0.8, color = red] (1.044, 0.937) +(-1.5, -0.5) arc (130: 465: 0.3) node[pos = 0.6, shift = {(0.35, -0.6)}, black] {\small$ \cW_C^{ \, \raisebox{0.6em}{\scriptsize$\frac{s'p}{\gcd(p,\ell)}$} }$};   
\end{scope}
\end{tikzpicture}
\caption{\label{fig: cylinder}%
Braiding of boundary line symmetry defects of the $\mathbb{Z}_L$ 1-form symmetry, and their cylindrical-surface counterparts in the bulk. Left: Symmetry defects on the boundary. Right: When the defects are pulled into the bulk, they become the boundaries of cylindrical bulk surface operators connected to the boundary. In this way the braiding of boundary lines is reproduced in the bulk.}
\end{figure}

On the other hand, if we try to extract a generic boundary line $\bigl( \wh W_C \bigr)^s$ from the boundary and move it into the bulk, it remains connected to the boundary through an attached cylindrical surface 
\be
\exp\biggl[ \frac{is}{L} \biggl(\, \int_\gamma q\, C + \int_{T_2} q\ell\, B - \int_{\hat\gamma} y\, \phi \biggr) \biggr] \,,
\ee
where $\gamma$ is a bulk line, $\hat\gamma$ is a boundary line, and $T_2$ is a cylindrical surface with $\partial T_2 = \gamma - \hat\gamma$. In the special case of $SU(N)_k$ where $p=1$ (and hence $q=L=N$, $y=N\ell$), these are all the operators $(\cW_C)^s$ deformed to end on a line of $\phi$ at the boundary. 
Now, a configuration of two cylinders attached to the boundary (similarly to the configuration of two semi-infinite cylinders discussed in the context of Lagrangian algebras) can indeed braid in four dimensions, as shown in Fig.~\ref{fig: cylinder}. 

Let us explore the relation between bulk and boundary lines more in detail. From (\ref{boundarybraiding}), the minimal exponent $s_*$ such that the boundary line $\bigl( \wh{W}_C \bigr){}^{s_*}$ has trivial braiding with all other topological boundary lines is
\be
s_* = \frac{L}{\gcd\bigl( L \,,\,  \frac{\ell}{\gcd(p,\ell)} \bigr) } \,.
\ee
This is one of the fractions appearing in (\ref{WCrelationbulkboundary}).
The trivial boundary line has exponent $s=L$ which is a multiple of $s_*$. On the other hand, the minimal exponent $s_\text{free}$ such that the boundary line $\bigl( \wh{W}_C \bigr){}^{s_\text{free}}$ is free to leave the boundary is%
\footnote{Indeed $s_\text{free}$ is the minimal positive integer such that the coefficient of $\phi$ in (\ref{boundary line operator}) is a multiple of $q$.}
\be
s_\text{free} = \frac{L\,q}{\gcd\bigl( L\,q, y \bigr)} = \frac{q}{\gcd\bigl( q, \frac{\ell}{\gcd(p,\ell)} \bigr)} = \frac{p}{\gcd(p, \ell)} \, s_* \,.
\ee
We see that $s_\text{free}$ is a multiple of $s_*$, and more precisely the boundary line $\bigl( \wh{W}_C \bigr){}^{s_\text{free}}$ coincides with the genuine bulk line $(W_C)^{N/g}[\gamma]$ discussed in \eqref{WCrelationbulkboundary}.
Thus we observe that for a topological boundary line of the $\bZ_L$ 1-form symmetry, in order to be able to freely dive into the bulk without an attached tube, it is not enough to have trivial braiding with all other lines of $\bZ_L$ (that would be a generic line generated by $\bigl( \wh{W}_C\bigr){}^{s_*}$): it has to satisfy a stronger constraint. This is because our Symmetry TFT also describes the $\bZ_p$ 0-form symmetry, and its $\bZ_{p/\gcd(p,\ell)}$ subgroup does not act on anything. This means that the corresponding symmetry defects are endable, and if we cut them open, their boundaries provide new topological lines. Only those lines of $\bZ_L$ that have trivial braiding also with the boundaries of the $\bZ_{p/\gcd(p,\ell)}$ endable symmetry defects%
\footnote{The 3d braiding between symmetry-defect lines of $\bZ_L^{[1]}$ and boundaries of endable symmetry defects of $\bZ_p^{[0]}$ is described by the first formula in \eqref{braiding in SymTFT}, with the substitution $\wh{W}_C = (\cW_C)^{p/\gcd(p,\ell)}$ and setting $m = gm'$ in order to select endable surfaces. Thus, if we restrict to mutually-transparent $\bZ_L^{[1]}$ lines by setting $r = \bigl( p / \gcd(p,\ell) \bigr) s_* r'$ and we require that the braiding trivializes for all $m'$, we find that $r'$ must be a multiple of $p/ \gcd(p,\ell)$.}
can freely dive into the bulk as genuine lines. Such boundary lines are precisely the ones generated by $\bigl( \wh{W}_C \bigr){}^{s_\text{free}}$.

The $\bZ_p$ 0-form symmetry and $\bZ_L$ 1-form symmetry realized by the symmetry defects on the topological boundary, or equivalently by their cylindrical avatars in the bulk, are precisely the symmetries of $\bigl( SU(N)/\bZ_p \bigr){}_k$ YM-CS theory, as reviewed in Appendix~\ref{app: SU(N)_k/Z_p 1-form sym}.


\subsection{Global variants with unfaithful symmetries}
\label{subsection: unfaith sym}

As we just saw, the global variant $\bigl( SU(N)/\bZ_p \bigr){}_k$ has $\bZ_p$ 0-form symmetry and $\bZ_L$ 1-form symmetry, with $L$ given in \eqref{def of L}.
This shows that, in contrast with what typically happens in theories with Abelian symmetries, the number of symmetry elements --- \ie{} of group elements --- is not the same across the various global variants. However, the total number of symmetry defects \emph{is} the same in all global variants once we include in the counting also the non-genuine boundaries of endable defects.%
\footnote{Since we are dealing with invertible symmetries, it is enough to count the number of topological operators. In the case of non-invertible symmetries there can be ``multiplicities'' in this counting, related to the quantum dimensions of non-invertible lines.}
This essentially holds by construction, since we specify the topological boundary conditions in terms of Lagrangian algebras in the bulk theory --- with the generalized definition of Lagrangian algebra: a Lagrangian algebra has a number of elements equal to the square root of the total number of topological operators of the bulk Symmetry TFT, and the symmetry defects of the boundary theory are the bulk operators modulo the Lagrangian algebra. 

From a three-dimensional boundary perspective, the reason for this atypical behavior is that, in most variants of YM-CS theory, a subgroup of the 0-form symmetry $\bZ_p$ does not act on anything. Indeed, as already discussed when analyzing the Symmetry TFT, when a symmetry acts trivially, its topological symmetry defects are endable also in the boundary.  Consequently, one could independently count both the symmetry defects and their boundaries. By employing this counting one can show, also from a purely  boundary point of view, that the number of symmetry defects is constant across all global variants of a given three-dimensional theory. Let us see how this works in the case of $\su(N)_k$ YM-CS theories.

\begin{table}
\centering
\renewcommand{\arraystretch}{1.4}
\begin{tabular}{|c||c|c|}
\hline
 & \hspace*{5mm} Electric 1-form \hspace*{5mm} & \hspace*{3mm} Magnetic 0-form \hspace*{3mm} \\[-0.7em]
 & or $\; \exp\bigl( i \!\int\! C \bigr)$ & or $\; \exp\bigl( i \!\int\! B \bigr)$ \\[0.2em] \hline\hline
\# of nontrivial bulk operators & $N$ & $N$ \\[0.3em] \hline
\# of genuine/non-endable bulk operators & $\gcd(N,k)$ & $\gcd(N,k)$ \\[0.3em] \hline
\# of elements in Lagrangian algebra & $p$ & $q=N/p$ \\[0.3em] \hline\hline
\# of boundary symmetry elements & $L = \dfrac{q}{p/\gcd(p,k)}$ & $p$ \\[0.7em] \hline
\raisebox{-0.1em}{\begin{tabular}{c} \# of faithfully-acting classes of \\[-1em] boundary symmetry elements \end{tabular}} & $L = \dfrac{q}{p/\gcd(p,k)}$ & $\gcd(p,k)$ \\[1em] \hline
\# of total boundary symmetry defects & \multicolumn{2}{c|}{ $L \, \times \, \gcd(p,k) \, \times \, \bigl( p/\gcd(p,k) \bigr)^2 = N$} \\[0.3em] \hline
\end{tabular}
\caption{\label{magictable}%
Summary of the cardinality of various sets of operators in the global variant $\bigl( SU(N)/\bZ_p \bigr){}_k$. In the first three rows we count the bulk topological operators of the Symmetry TFT constructed in (\ref{def of W_B}), (\ref{nongenl}) and above (\ref{def of g}), while the (generalized) Lagrangian algebra was defined below (\ref{spin in SymTFT}). In the last three rows we count the symmetry defects of $\bZ_L^{[1]}$ and $\bZ_p^{[0]}$ in the boundary theory, keeping into account that the defects of the $\bZ_{p/\gcd(p,k)}^{[0]}$ subgroup do not act on anything, are endable, and thus their boundaries should be independently counted.}
\end{table}

The theory $SU(N)_k$ has a 1-form symmetry $\bZ_N^{[1]}$. Among its lines, the ones that do not braid with any other line form a subgroup $\bZ_g^{[1]}$ generated by $U_{N/g}$ (where $g = \gcd(N,k)$). These lines are special because they can terminate on a point, thus defining twisted sectors. All other lines cannot be broken, because they carry some charge under $\bZ_N^{[1]}$. Let us focus on the subgroup $\bZ_p^{[1]} \subset \bZ_N^{[1]}$ generated by the line $U_{N/p}$. In this subgroup, the endable lines that do not braid with any other line of $\bZ_N^{[1]}$ form a subgroup $\bZ_{\gcd(p,k)}^{[1]}$ generated by $U_{N/\gcd(p,k)}$. Going to the quotient, there are $p/\gcd(p,k)$ classes of lines in $\bZ_p^{[1]}/\bZ_{\gcd(p,k)}^{[1]}$ such that the trivial class contains the endable lines while the other classes contain the unbreakable lines.

When we gauge $\bZ_p^{[1]}$ in order to obtain the variant $\bigl( SU(N)/\bZ_p \bigr){}_k$, a dual 0-form symmetry $\bZ_p^{[0]}$ arises. Local operators charged under $\bZ_p^{[0]}$ are monopole operators, which before the gauging were endpoints of the lines of $\bZ_p^{[1]}$ to be gauged. As discussed above, there is an obstruction to the existence of gauge-invariant monopole operators with charge $n \in \bZ_p^{[0]}$, because the corresponding line of $\bZ_p^{[1]}$ before gauging has to be in the trivial class. In particular, there exist monopole operators only when the charge $n$ is a multiple of $p/\gcd(p,k)$. This implies that the symmetry $\bZ_p^{[0]}$ does not act faithfully: its subgroup $\bZ_{p/\gcd(p,k)}^{[0]}$ does not act on anything.

We conclude that the surface symmetry defects of the subgroup $\bZ_{p/\gcd(p,k)}^{[0]} \subset \bZ_p^{[0]}$ are endable, and hence we should count their boundaries as well. All in all we hence find
\be
\text{number of symmetry defects} = L \times p \times \frac{p}{\gcd(p,k)} = N \;.
\ee
The first factor comes from $\bZ_L^{[1]}$, the second factor from $\bZ_p^{[0]}$, and the third factor from the boundaries of the endable surfaces of $\bZ_{p/\gcd(p,k)}^{[0]} \subset \bZ_p^{[0]}$. For convenience, Table~\ref{magictable} summarizes the cardinality of the various sets of objects discussed in this section.


\section{Symmetry TFT from holography}
\label{section: holographic symTFT}

As we are going to show in the following, the Symmetry TFT for $\su(N)_k$ YM-CS theories, including the corresponding symmetry defects and charged operators, can also be derived from a (fully top-down) holographic setup in type IIB string theory.


\subsection{Holographic model}
\label{subsection: holo model}

Three-dimensional $SU(N)$ YM theory (we are not paying attention to the global structure of the gauge group, for the time being) can be engineered in type IIB string theory placing $N$ D3-branes along the directions $x^0,\dots,x^3$, with $x^3$ compact with period $2\pi/M_{KK}$ \cite{Witten:1998zw}. Imposing anti-periodic boundary conditions for fermions and periodic for bosons on $x^3$, supersymmetry is explicitly broken and a mass of order $M_{KK}$ for both gaugini (at tree level) and scalar fields (at one-loop) is generated, while the three-dimensional gauge bosons remain massless. Therefore, the four-dimensional $\mathcal{N}=4$ SYM theory reduces at low energies (below $M_{KK}$) to pure $SU(N)$ YM theory in three dimensions, with gauge coupling $g_{\rm YM}^2=g_s M_{KK}$, $g_s$ being the string coupling constant.
The dual type IIB supergravity background contains the metric, the dilaton $\phi$ and the Ramond-Ramond (RR) five-form $F_5$, and reads (in Lorentzian signature, which we will adopt in this section):
\bea
\label{background}
ds^2 &= \frac{r^2}{L^2} \Bigl( \eta_{\mu\nu} dx^\mu dx^\nu + f(r) (dx^3)^2 \Bigr) + \frac{L^2}{r^2 f(r)} dr^2 + L^2 ds^2_{S^5} \,, \\
e^\phi &= g_s \,, \qquad\qquad \frac{1}{(2\pi l_s)^4 }\int_{S^5} F_5 = -N \,,
\eea
where $ds_{S^5}^2$ is the metric on a unit five-sphere, $\mu,\nu=0,1,2$ and
\be
L^4=4\pi g_s N l_s^4 \,, \qquad\qquad f(r)=1 - \biggl( \frac{r_0}{r} \biggr)^4 \,, \qquad\qquad r_0 = \frac{M_{KK} L^2}{2} \,,
\ee
with $l_s$ the string length. We take $l_s=1$ from now on. The geometry is given by a flat $\mathbb{R}^{2,1}$, a constant $S^5$ and a cigar-shaped $(r,x^3)$ subspace, where the holographic coordinate $r$ goes from $r=\infty$ at the holographic boundary to $r=r_0$ at the tip of the cigar. Here the geometry smoothly ends, thus giving rise to a mass gap (glueballs get a mass of order $M_{KK}$) and to an area law for the fundamental Wilson loop (the holographic string tension scales as $(\Lambda M^3_{KK})^{1/2}$, where $\Lambda=g_{\rm YM}^2 N$ is the strong coupling scale). This implies that the theory confines, and its 1-form global symmetry is unbroken. 

In three spacetime dimensions a Chern--Simons term can be added to a gauge theory. This can be implemented in the holographic setup following \cite{Fujita:2009kw}. The D3-brane theory admits a coupling with the RR axion $C_0$ 
\be
\label{C0coupl}
S_{C_0} = \frac{1}{4\pi} \int_\text{D3} C_0 \Tr (f \wedge f) = \frac{1}{4\pi} \int_{\mathbb{R}^{2,1} \times S^1} \Omega_3(a) \wedge F_1 \,,
\ee
where $a$ is the gauge field on the D3-branes, $f$ is its field strength, and $\Omega_3(a)$ is the CS form in three dimensions. Assuming that $a$ does not depend on $x^3$ nor it has components along $S^1$, and choosing a background value for the RR flux $F_1=dC_0$ such that
\be
\label{rrpurecs}
\int_{S^1} F_1  = k \,,
\ee
a CS term at level $k$ (which we take to be positive without loss of generality) for the D3 gauge field $a$ is produced:
\be
\label{C0CS}
S_{C_0} = \frac{k}{4\pi} \int_{\mathbb{R}^{2,1}} \Omega_3(a) \,.
\ee
Since we are neglecting the backreaction of the axion field on the D3 background, strictly speaking this description is only valid when $k \ll N$. Note that this implies that the strong coupling scale $\Lambda=g_{\rm YM}^2N$ is much larger than the tree-level gluon mass $m_g=g_{\rm YM}^2k$. Hence, this theory is really the full YM-CS theory, and gluons cannot be integrated out at weak coupling.

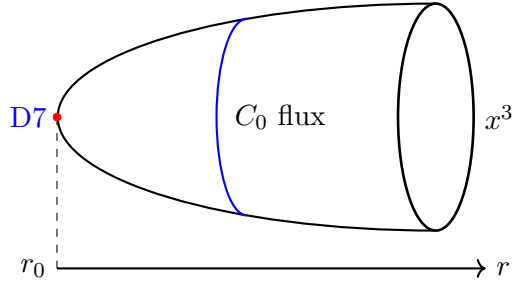
\begin{figure}[t]
\centering
\begin{tikzpicture}
	\draw[thick] (4.34, -1.5) +(90: 5cm and 1.5cm) arc (90: 270: 5cm and 1.5cm);
	\draw[line width = 1] (4.34, -1.5) ellipse [x radius = 0.5, y radius = 1.5] node[right = 0.5 cm] {$x^3$};
	\draw[thick, blue] (4.34, -1.5) +(120: 5cm and 1.5cm) arc (90: 270: 0.4cm and 1.299cm) node[pos = 0.5, right = 0.1 cm] {\color{black}$C_0$ flux};    
	\draw[thick,->] (-0.67, -3.5) node[left] {$r_0$} -- (5, -3.5) node[right] {$r$} ;
	\draw[dashed] (-0.67, -1.5) -- (-0.67, -3.5);
	\filldraw [red] (-0.67, -1.5) circle [radius = 0.05] node[left, black] {\color{blue}D7} ;
\end{tikzpicture}
\caption{\label{CSsig}%
The D7-branes which act as a source for the RR $C_0$ potential around $S^1$ are located at the tip of the cigar geometry.}
\end{figure}

At strong coupling the stack of D3-branes is replaced by the cigar geometry. This is topologically a disk, so the presence of a non-vanishing flux on the contractible $S^1$ needs to be supported by a magnetic source for $C_0$. This is provided by $k$ D7-branes wrapped on $S^5$, that are pointlike on the cigar and extend in the three dimensions of Minkowski spacetime. To minimize their energy density, the D7-branes are located at the tip of the cigar, where the $x^3$ circle shrinks to a point, see Fig.~\ref{CSsig}. In this situation, the worldsheet of a string which is attached to a fundamental Wilson loop at the boundary can end on the D7-branes at the tip of the cigar. This configuration is energetically favored, as it was explicitly computed in \cite{Fujita:2016gmu}, and it signals a perimeter law for the Wilson loop. This shows that in the presence of a Chern--Simons term the theory does not confine. Equivalently, the electric 1-form symmetry is spontaneously broken.

At energies below the $S^5$ inverse radius, the D7-brane theory reduces to a three-dimensional gauge theory with gauge group $U(k)$. Note that the D7-branes are stuck at the tip, in their vicinity the ambient space is locally flat, and the (low-energy effective) theory described by the open strings ending on them is indeed $U(k)$. 

The presence of a background 5-form flux induces a Chern--Simons term at level $-N$ from the corresponding Wess--Zumino term in the D7-brane action
\be
\label{SC4}
S_{C_4} = \frac{1}{2 (2\pi)^5} \int_\text{D7} C_4 \wedge \Tr ( \cF \wedge \cF) =
\frac{1}{2 (2\pi)^5} \int_{\mathbb{R}^{2,1} \times S^5} \Omega_3(b) \wedge F_5 = - \frac{N}{4\pi}\int_{\mathbb{R}^{2,1}} \Omega_3(b) \,,
\ee
where $b$ is the worldvolume gauge field on the D7-branes and $\cF$ is its field strength. Note that, because the level $N$ is much larger than the rank $k$, the gluons of this theory can be integrated out at weak coupling. Thus, at low energies all excitations on the D7-branes decouple and we are left with a pure $U(k)_{-N}$ CS theory.

All in all, we started from an $SU(N)_k$ YM-CS theory with large $N$ and fixed $k$ in the UV, and we ended up with a $U(k)_{-N}$ pure CS theory in the IR. By level/rank duality, this is equivalent to an $SU(N)_k$ pure CS theory. Thus, this model describes the RG flow from a YM-CS theory to a pure CS theory.%
\footnote{Strictly speaking, this would require the presence of a regime where the UV theory is perturbatively YM-CS. In the holographic regime, however, the theory in the UV is never decoupled from the tower of KK modes, as in many other holographic RG-flow scenarios. We thus need to assume that the holographic theory at hand and $SU(N)_k$ YM-CS are in the same universality class.}
We will come back to this RG flow in Section~\ref{section: holographic RG}.

Finally, let us clarify an issue related to the global structure of the gauge group. When we consider the theory on the D3-branes, the gauge field $a$ takes values in the Lie algebra $\mathfrak{u}(N)$. In order to get a theory with an $SU(N)$ gauge group, we need to impose Dirichlet boundary conditions on the NSNS 2-form $B_2$ \cite{Witten:1998wy}. This implies that the RR 2-form $C_2$ is free to vary on the boundary. From the point of view of the D3-branes, whose action contains the term
\be
S_{C_2} \sim \int_\text{D3} C_2 \wedge \Tr(f) \sim \int_{\mathbb{R}^{2,1}} C \wedge \Tr(f) \,, 
\ee
the field $C$ --- which is the dimensional reduction of $C_2$ on the circle --- acts on the boundary as a Lagrange multiplier that sets $\Tr(f) = 0$ once we path-integrate over it, effectively reducing the $U(N)$ gauge field to its traceless $SU(N)$ part \cite{Argurio:2022vtz}. Different boundary conditions are possible, and they correspond to different global variants of $SU(N)$, in a similar vein as in Section \ref{section: symTFT in QFT}.


\subsection{Topological action from type IIB supergravity}
\label{subsection: action from holography}

Let us now derive the four-dimensional topological sector of the supergravity theory discussed above. As we are going to show, such sector captures the global symmetries of the boundary YM-CS theory and provides the action of the Symmetry TFT. To do that, we compactify the ten-dimensional equations of motion on $S^1$ and $S^5$, and then write down a 4d action which reproduces them.

Let us start from the bosonic part of the action of type IIB supergravity in the string frame, neglecting its gravity and dilaton parts:
\be
S_\text{IIB} \,\supset\, -\frac{1}{2(2\pi)^7} \int \biggl( e^{-2\phi} \, H_3 \wedge \star\,H_3 + F_1 \wedge \star\,F_1 + \widetilde{F}_3 \wedge \star\,\widetilde{F}_3 + \frac{1}{2} \, \widetilde{F}_5 \wedge \star\,\widetilde{F}_5 + C_4 \wedge H_3 \wedge F_3 \biggr) , 
\ee
where $H_3 = dB_2$, $F_{p+1} = dC_p$ and
\be
\widetilde{F}_3 = F_3 - C_0 H_3 \,, \qquad \widetilde{F}_5 = F_5 + \frac{1}{2} \, B_2 \wedge F_3 - \frac{1}{2} \, C_2 \wedge H_3 \,.    
\ee
The self-duality condition $\widetilde{F}_5 = \star\,\widetilde{F}_5$ has to be imposed after the equations of motion are derived. In the supergravity background of interest, we have
\be
\label{background forms}
F_1 = k \, \omega_1 \,, \qquad\qquad F_5 = -(2\pi)^4 N \omega_5 \,, 
\ee
where $\omega_i$ are the unit volume forms on $S^1$ and $S^5$. The equation of motion for $C_0$ is
\be
d \star F_1 + H_3 \wedge \star\,\widetilde{F}_3 = 0 \,,
\ee
and it is automatically satisfied by the background in \eqref{background forms}. This means that if $B_2$ and $C_2$ fluctuate, they are forced to satisfy
\be
\label{C0equation}
H_3 \wedge \star\,\widetilde{F}_3 = 0 \,.   
\ee
The equation of motion for $C_4$ is
\be
\label{c4equation}
d\star \widetilde{F}_5 - H_3 \wedge F_3 = 0 \,,
\ee
which can be easily shown to be equivalent to the Bianchi identity $dF_5=0$ once the self-duality condition is imposed. The most relevant equations for us are the ones for $C_2$ and $B_2$, which read 
\be
\label{eomIIB}
d\star \widetilde{F}_3 + H_3 \wedge \widetilde{F}_5 = 0 \,,\qquad\qquad
-g_s^{-2} \, d\star H_3 + d \bigl( C_0 \star \widetilde{F}_3 \bigr) + F_3 \wedge \widetilde{F}_5 = 0 \,.
\ee
Assuming that all fields but $F_5$ do not have components along $S^5$, the first equation is solved by
\be
\label{eqtildeF3}
\star\,\widetilde{F}_3 = dC_6 - B_2 \wedge F_5 \,.
\ee
Taking $B_2$ not to have components along $S^1$,%
\footnote{An $S^1$ component of $B_2$ would describe the background 1-form gauge field of the electric 0-form symmetry associated to the circle compactification. We do not include this component because we are not interested in such a symmetry, which is not present in the low-energy field theory.}
we have
\be
\star\,H_3 = (2\pi)^4 \, h(r) \, (\star_4\,H_3) \wedge \omega_1 \wedge \omega_5 \,,   
\ee
where $h(r)$ is a function of the radial coordinate only (whose precise expression is not needed) and $\star_4$ is the 4d Hodge star. Making the following Ansatz for the dimensional reductions of $C_6$ and $C_2$,
\be
\label{AnsC6C2}
C_6 = (2\pi)^5 \, \widetilde{C} \wedge \omega_5 \,, \qquad\qquad C_2 = 2\pi \, C \wedge \omega_1 \,,
\ee
and using \eqref{background forms}, the second equation in \eqref{eomIIB} becomes 
\be
\Bigl[ \Bigl( -g_s^{-2} \, d \bigl( h(r) \star_4 dB \bigr) + k \, d\widetilde{C} + k N \, B - N \, dC \Bigr) \wedge \omega_1 + N \, C_0 \, dB \Bigr] \wedge \omega_5 = 0  \,,
\ee
where we normalized the $B_2$ field as $B_2 = 2\pi B$. Since $dB$ does not have any component along $S^1$, this is equivalent to the two equations
\be
\label{reduced sugra eqs}
k \, d\widetilde{C} + k N \, B - N \, dC = 0 \,,\qquad\qquad N \, dB = 0 \,.
\ee
Note that the equation \eqref{C0equation} is also satisfied. The equations in \eqref{reduced sugra eqs} can arise as the equations of motion of the following four-dimensional action:
\be
S_\text{4d} = \int \biggl( - \frac{N}{2\pi} \, B \wedge dC + \frac{k}{2\pi} \, B \wedge d\widetilde{C} + \frac{Nk}{4\pi} \, B \wedge B \biggr) \,,
\ee
where $C$ and $\widetilde{C}$ are not independent, but rather are related by the ten-dimensional Hodge duality relation \eqref{eqtildeF3} which becomes
\be
\label{reduced hodge star}
2\pi \star \bigl( dC \wedge \omega_1 \bigr) = (2\pi)^5 \bigl( d\widetilde{C}+  N \, B \bigr) \wedge \omega_5 \,.
\ee
Using the large $r$ dependence of the background metric \eqref{background}, one can translate this relation to one that involves the 4d Hodge star:
\be
d \widetilde{C} + N \, B \sim \frac{1}{r} \star_4 dC \,.
\ee
This means that for large values of $r$ (\ie{} near the holographic boundary) we have $d\widetilde{C}=-NB$. All in all, close to the boundary the bulk theory is dominated by the following action:
\be
\label{SymTFT1}
S_\text{4d} = - \int \biggl( \frac{N}{2\pi} \, B \wedge dC + \frac{Nk}{4\pi} \, B \wedge B \biggr) \,,   
\ee
which is (after a Wick rotation to Euclidean signature) the Symmetry TFT action \eqref{eq: symTFT}, including the quadratic term in $B$ which captures the 't~Hooft anomaly of the $\mathbb{Z}_N$ 1-form symmetry (recall that, by construction, in our top-down holographic setup we deal with spin manifolds, so $\ell =k$ here).


\subsection{Symmetry defects from holography}

Let us now see how the topological operators discussed in Section~\ref{section: symTFT in QFT} are described holographically. Let us start with the $SU(N)$ global form. Under the $\mathbb{Z}_N$ 1-form symmetry, the minimally charged operator is a fundamental Wilson line supported on a curve $\gamma'$. This is described by a fundamental string whose two-dimensional worldsheet $\Sigma'_2$ extends in the bulk and ends on $\gamma'$ at the $r = \infty$ boundary of the geometry \eqref{background}, as in ordinary holographic constructions \cite{Maldacena:1998im, Rey:1998ik}. Fundamental strings couple to the NSNS two-form $B_2=2\pi B$ as
\be
S_\text{F1} = \int_{\Sigma'_2} B \,,  
\ee
where $\partial \Sigma'_2=\gamma'$. The fundamental Wilson line hence reads $W_B[\Sigma'_2] = \exp \bigl( i \int_{\Sigma'_2} B \bigr)$. The operator that measures the charge of such a Wilson line is a line operator of $C$, which corresponds to a D1-brane with one direction wrapped on the circle and the other direction extended along the boundary. Let us briefly recall why this is the case. Introducing a fundamental Wilson line in the effective 4d topological action \eqref{SymTFT1} leads to 
\be
S_\text{4d} + S_\text{F1} = - \int \biggl( \frac{N}{2\pi} B \wedge dC + \frac{Nk}{4\pi} B \wedge B \biggr) + \int B \wedge \delta^{(2)}(\Sigma'_2) \,,
\ee
where $\delta^{(2)}$ localizes the four-dimensional integral on the worldsheet. The equation of motion for $B$ and its restriction to the 3d boundary (where $B=0$ due to the Dirichlet boundary conditions) read, respectively:
\be
\label{B eom bulk/boundary}
\frac{N}{2\pi} \, dC + \frac{Nk}{2\pi} \, B = \delta^{(2)}(\Sigma'_2) \,, \qquad\qquad \frac{N}{2\pi} \, dC \Big|_{r=\infty} = \delta^{(2)}(\gamma') \,.
\ee
The latter is exactly the violation of the conservation of the electric 1-form symmetry current $J_e$ in the presence of an electrically charged Wilson line, once we identify
\be
\star_3\,J_e = \frac{N}{2\pi} \, C \Big|_{r=\infty} \,.     
\ee
We need $C|_{r=\infty}$ to be a freely-varying field, namely Neumann boundary conditions for $C$. The symmetry defect of the electric $\mathbb{Z}_N$ 1-form symmetry, supported on a line $\gamma$ at the boundary $r=\infty$, is thus given by the GW operator
\be
U_{\alpha = \frac{2\pi p}{N}}[\gamma] = \exp \biggl( i \alpha \int_\gamma \star_3\,J_e \biggr) = \exp \biggl( ip \int_\gamma C \biggr)
\ee
with integer $p\sim p+N$. This is nothing but the (exponential of the) WZ action of $p$ coincident D1-branes wrapped on $S^1$ and supported on a line $\gamma$ on the 3d boundary. Indeed, using the Ansatz \eqref{AnsC6C2}, the WZ action for a D1-brane on the boundary (where $B=0$) reads
\be
\label{D1 WZ action}
S_\text{D1} = \frac{1}{2\pi}\int_{\gamma \times S^1} \Bigl( C_2 - 2\pi C_0 B \Bigr) = \int_\gamma C \,.
\ee
Hence, D1-branes at the boundary are associated with the $\bZ_N$ topological symmetry defects of the 3d $SU(N)_k$ YM-CS theory. 

\begin{figure}[t]
\centering
\begin{tikzpicture}[scale=1]
	\fill [green] (-0.1, 0) -- (0.9, 0.7) -- (1, 0.7) -- (0, 0) -- cycle;
	\shade [bottom color = red, top color = red!75!green] (0,0) arc [start angle = 0, end angle = 90, x radius = 1, y radius = 2] -- (-1,0) -- cycle;
	\shade [top color = red!75!green, bottom color = red!25!green] (-1,2) arc [start angle = 90, end angle = 270, x radius = 1, y radius = 2] -- cycle;
	\shade [bottom color = red!25!green, top color = green] (0,0) arc [start angle = 0, end angle = -90, x radius = 1, y radius = 2] -- (-1,0) -- cycle;
	\fill [white] (-0.1, 0) arc [start angle = 0, end angle = 360, x radius = 0.9, y radius = 1.9];
	\draw [->, thick] (-0.3, 0) node [left] {\small$x_3$} arc [start angle = 0, end angle = 25, x radius = 0.8, y radius = 1.8];
	\node at (0.2, 0.8) {\small{}D1};
	\node [right] at (-0.1, -0.5) {\small{}D1/$k$ F1};
	\node at (1.4, 0.7) {\small$k$ F1};
\begin{scope}[shift={(7, -1)}]
    \draw [dashed, red, line width=1.5] (0,0) +(0:1.81 and 0.7) arc [start angle = 0, end angle = 180, x radius = 1.81, y radius = 0.7];
    \fill[color = white, bottom color = white!30!green, top color = white, opacity=0.5] (-1.81, 0) -- (-1.81, 2) -- (1.81, 2) -- (1.81, 0);
    \fill[color=white!30!green, opacity=0.5] (0,0) +(0:1.81 and 0.7) arc [start angle = 0, end angle = -180, x radius = 1.81, y radius = 0.7];
    \draw (-1.81, 0) -- (-1.81, 2);
    \draw (1.81, 0) -- (1.81, 2) node [below left, color=black] {$k$ F1};
    \draw[red, line width=1.5] (0,0) +(0:1.81 and 0.7) arc [start angle = 0, end angle = -180, x radius = 1.81, y radius = 0.7] node [below left, color=black] {D1};
\end{scope}
\end{tikzpicture}
\caption{The stringy bound states describing bulk topological operators of $SU(N)_k$ YM-CS theory. Left: a D1-brane (red) wrapping the circle direction $x_3$ by Witten effect develops $k$ units of F1-charge, so that $k$ F1-strings (green) are emitted in an orthogonal direction. Right: a D1-brane along a curve $\gamma$ (and wrapping $x_3$, not drawn) must live at the boundary of $k$ F1-strings.
\label{fig: D1/F1 brane}}
\end{figure}
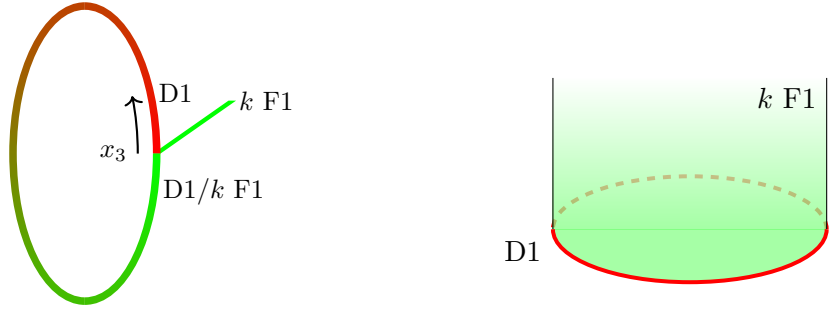

If $k=0$, a D1-brane describing a symmetry defect of the boundary YM-CS theory, whose WZ action is \eqref{D1 WZ action}, can be moved into the bulk and the operator $W_C[\gamma]=\exp \bigl( i \int_\gamma C \bigr)$ is well defined even if $\gamma$ does not lie on the boundary. If $k\neq 0$, instead, this is not the case. We learned, following the discussion  in Section~\ref{section: symTFT in QFT}, that when $k \not =0$ gauge invariance requires the symmetry defect $W_C[\gamma]$ to be attached to a surface $D_2$ whose boundary is $\partial D_2=\gamma$:
\be
\label{stringy charges}
W_C[\partial D_2] \; (W_B)^k[D_2] = \exp \biggl( i \int_{\partial D_2} C + ik \int_{D_2} B \biggr) \,. 
\ee
This has a natural explanation in the holographic construction. When $k\neq 0$, along the circle $S^1$ there is a varying background for $C_0$ that solves \eqref{rrpurecs}, given by $2\pi C_0 = k \varphi$ where $\varphi = x_3 M_{KK}$ is the angular coordinate along the $S^1$. Since the action $\eqref{D1 WZ action}$ wraps the circle, as we move from $\varphi = 0$ to $\varphi=2\pi$ an effective F1-charge builds up on the worldvolume of the D1-brane (due to the Witten effect \cite{Witten:1979ey}), so that once we reach $\varphi = 2\pi$ the brane looks like a D1/$k$~F1 bound state. This point however has to be glued to the point $\varphi = 0$, where there is just a D1-brane, therefore $k$ F1-strings have to come out along an orthogonal direction.%
\footnote{Alternatively and more simply, the coupling $\int\! dC_0 \wedge A$ on the D1-brane involving the worldvolume gauge field $A$ creates $k$ units of charge, and in order to avoid a tadpole, $k$ F1-strings need to terminate on the D1. We thank Oren Bergman for pointing this out to us.}
This is pictured in Fig.~\ref{fig: D1/F1 brane} left. The net result is that a D1-brane wrapping $S^1$ and placed along a curve $\gamma$ must live at the boundary of $k$ F1-strings with worldsheet $D_2$ such that $\partial D_2 = \gamma$, as shown in Fig.~\ref{fig: D1/F1 brane} right (where the circle direction $x_3$ is not drawn). This system reproduces the operators \eqref{stringy charges}, which are nothing but \eqref{nongenl} (recall that $\ell =k$ here).

We conclude that our holographic model exactly reproduces the Symmetry TFT and all symmetry defects and charged operators of $SU(N)_k$ YM-CS theory, including also endable operators, which, as argued by the field theory analysis of Section~\ref{section: symTFT in QFT}, are crucial to make sense of the Symmetry TFT. The description of other global variants are obtained imposing appropriate boundary conditions for the $B$ and $C$ fields at $r = \infty$. The discussion follows verbatim the one in Section~\ref{section: symTFT in QFT} and we do not repeat it here. 

It would seem that this agreement holds as a general rule, since the standard results of \cite{Witten:1998wy}. However, in holography the $(d+1)$-dimensional bulk is not topological, not even in the near-boundary region where --- as already emphasized --- the correspondence with the Symmetry TFT description is expected to hold. There can be instances where holography, while still capturing the topological aspects of the dual gauge theory, does not reduce to the Symmetry TFT. Natural candidates are gauge theories with $U(1)$ factors, whose holographic description requires dynamical boundary degrees of freedom and where bulk field boundary conditions are not purely topological. This will be addressed in \cite{noi}.


\section{Holographic RG flow from YM-CS to pure CS}
\label{section: holographic RG}

In this section we want to show how the bulk description encodes the RG flow that takes place in the 3d theory. In field theory terms, what we are describing is an RG flow from a YM-CS theory in the UV to a pure CS theory in the deep IR.
This means, in particular, that some line operators that are not topological in the UV should become topological along the flow. Some other line operators, on the other hand, are expected to disappear from the IR theory.

In YM-CS theory Wilson lines are not topological, and they are labeled by irreducible representations of the gauge group $SU(N)$, which can be described in terms of Young diagrams. The generators of the $\mathbb{Z}_N$ 1-form symmetry are instead topological lines.
When linking the Wilson lines, they measure their $\mathbb{Z}_N$ charge, which is given by the $N$-ality of the representation. Hence, in the UV there is a clear distinction between symmetry defects and the objects which they act on. As previously discussed, when the CS level $k \not =0$ the $\mathbb{Z}_N$ 1-form symmetry is anomalous and the topological lines are themselves charged, with charge multiple of $k$.

In the deep IR, \ie{} at energies well below the strong coupling scale $\Lambda = Ng_\mathrm{YM}^2$,
it is widely believed that the theory flows to a pure $SU(N)_k$ CS theory.%
\footnote{Remarkably, this is instead a prediction of the holographic model we are using. Indeed, on spin manifolds, we have that $U(k)_{-N} \leftrightarrow SU(N)_k$ as pure CS theories, by level/rank duality.}
In pure CS theory, there is only one kind of lines, the Wilson lines, and they are all topological. For $k \not =0$ Wilson lines are still labeled by irreducible representations of $SU(N)$, but only representations corresponding to Young diagrams with at most $k$ columns lead to non-vanishing Wilson lines. The $N$ lines generating the $\mathbb{Z}_N$ 1-form symmetry are the subset consisting of the Abelian lines, corresponding to rectangular Young diagrams with $m$ rows of $k$ boxes, with $m=0,\dots, N-1$. What distinguishes the Wilson lines that generate the  $\mathbb{Z}_N$ 1-form symmetry in the IR from all others is that they fuse according to the $\mathbb{Z}_N$ group law. Yet, also all other Wilson lines are topological and can be considered as 1-form symmetry defects. The symmetry generated by the full set of Wilson lines is the non-invertible 1-form symmetry of $SU(N)_k$ CS theory \cite{Witten:1988hf}.

Remarkably, all these aspects can be seen from the dual holographic perspective. 

We have already discussed Wilson lines in our setup. For gauge group $SU(N)$ they correspond to fundamental strings ending on the boundary, in agreement with the fact that we impose Dirichlet boundary conditions on the NSNS field $B_2$, which fundamental strings couple to. Let us start focusing on Wilson lines in the fundamental representation. If we take a line along the 3d boundary, then the worldsheet of the F1-string is an open 2d surface ``hanging" in the bulk along the radial direction, and whose boundary is the line at the boundary. For a line that is a small loop, the 2d surface will have the topology of a disk, and it will reach a minimal radius that is still well within the AdS region of the background. Hence, the Wilson loop one-point function will have a scaling typical of its UV completion, namely 4d conformal $\mathcal{N}=4$ SYM, for instance the potential between the probe quarks is Coulomb-like.

When the loop at the boundary grows, the minimal radius of the F1-string worldsheet goes deeper into the bulk, until it reaches the tip of the bulk geometry. This corresponds to a Wilson loop whose size is larger than the inverse of the strong-coupling scale $\Lambda$.
When the F1-string reaches the tip, it can end on the D7-branes that are located there, and which are parallel to the boundary (we take here the bulk to be the four-dimensional manifold $\mathcal{M}_4$ we get after the reduction on $S^5$ and $S^1$, so that D7-branes are effectively three-dimensional objects). Hence, the worldsheet can open there, so that it has now the topology of a cylinder, with the loop on the D7-branes being a sort of IR image of the loop at the boundary.\footnote{As already noticed in Section~\ref{subsection: holo model}, the expectation value of the Wilson loop satisfies a perimeter law typical of a deconfined phase with a spontaneously broken electric 1-form symmetry. Indeed, the potential between probe quarks is constant.}
For any $k\neq 0$, this configuration is actually energetically favored with respect to the disk topology, which for a large Wilson loop would describe instead an area law and a confining phase \cite{Fujita:2016gmu}. We would like to further argue that the dual bulk description gives evidence that the Wilson loop is in fact topological in the deep IR. To see this, notice that the boundary conditions at the two ends of the cylinder stretched between the boundary at $r=\infty$ and the D7-branes at the tip are actually different. As usual for holographic Wilson loops, the conditions at the boundary of $\mathcal{M}_4$ are of Dirichlet type \cite{Drukker:1999zq}. At the other end of the worldsheet the boundary conditions on the D7-branes are instead of Neumann type, from the very definition of what D-branes are. This is an indication that the line living on the D7-branes is topological. Indeed, consider the end of a string on the D7-branes as a probe quark in the theory defined on the worldvolume of the D7-branes. This is charged under the gauge group of the latter. But,  as we derived earlier, the gauge theory on the D7-branes is nothing but a $U(k)_{-N}$ pure CS theory (the YM term is suppressed already in the perturbative regime, because $N \gg k$ and gluons get a mass and can be integrated out semiclassically). Hence probe quarks have worldlines that are topological Wilson lines. We then establish that fundamental Wilson lines of the $SU(N)_k$ YM-CS theory become topological in the IR by manifesting themselves as the fundamental Wilson lines of the $U(k)_{-N}$ level/rank dual pure CS theory, which are obviously topological.

The above discussion can be extended to Wilson loops in other representations of $SU(N)$. Some of these operators should also become topological along the RG flow. These include the operators that are topological in the UV, \ie{} the symmetry defects (D1/$k$~F1 bound states in the holographic setup) which in the IR are identified with the Wilson lines in the corresponding representation. In fact, one expects to have the possibility of creating networks of surfaces in 4d that provide the putative Symmetry TFT for the full non-invertible 1-form symmetry of the pure CS theory in the IR.

For 4d ${\cal N}=4$ SYM a precise recipe has been developed to describe line operators in arbitrary representations as states of type IIB string theory in AdS$_5 \times S^5$  \cite{Drukker:2005kx, Yamaguchi:2006tq, Gomis:2006sb, Gomis:2006im}. Wilson line operators are embodied by either D3- or D5-branes with some units of fundamental string flux dissolved on their worldvolume and which end on lines at the boundary of AdS. The units of F1-string flux equal the number of boxes for each given column (D5-brane) or row (D3-brane) of the corresponding Young diagram. So, any given Young diagram can be described in terms of bound states of D5- or D3-branes, whose solutions as defects in AdS$_5 \times S^5$ have also been found.  

It would be very interesting to find analogous solutions in our setup, which is in one lower dimension and, most importantly, with broken supersymmetry. This is beyond the scope of the present paper, but we do not expect qualitative changes to what mostly concerns us here. Indeed one can think of configurations with a number of parallel F1-strings which equals the number of boxes of the desired Young diagram. This holographically represents a Wilson loop in a reducible representation, that contains the irreducible representations  corresponding to the Young diagrams of interest. The argument for their topological nature once the strings open up on the D7-branes at the tip goes verbatim as for the F1-string describing a Wilson loop in the fundamental representation.
However, on the D7-branes there are topological lines only in representations allowed for pure $U(k)_{-N}$ CS theory, which are the same as those of the level/rank dual $SU(N)_k$ CS theory. In particular, they can have at most $N$ rows and $k$ columns in terms of Young diagrams. Hence, only string configurations which are compatible with those representations can end topologically on the D7-branes at the tip. The other larger representations will just lie at the tip, and the corresponding Wilson loops will display an area law. In other words, they vanish in the deep IR, \ie{} they are not operators of the IR pure CS theory.

We then see how the holographic model of \cite{Witten:1998zw, Fujita:2009kw}, which describes three-dimensional $SU(N)_k$ YM-CS theory at large $N$ (or, more precisely, is believed to belong to the same universality class), correctly captures the way topological and non-topological operators rearrange or disappear along the RG flow from the YM-CS regime in the UV to the CS regime in the IR.


\section*{Acknowledgements}

We are grateful to Andrea Antinucci, Fabio Apruzzi, Oren Bergman, Christian Copetti, Rajath Radhakrishnan, Giovanni Rizi, and Luigi Tizzano for useful discussions and correspondence. R.A. is a Research Director of the F.R.S.-FNRS (Belgium). The research of R.A. and G.G. is supported by IISN-Belgium (convention 4.4503.15) and through an ARC advanced project. 
F.B. is supported by the ERC-COG grant NP-QFT No.~864583 ``Non-perturbative dynamics of quantum fields: from new deconfined phases of matter to quantum black holes'', by the MUR-FARE grant EmGrav No.~R20E8NR3HX ``The Emergence of Quantum Gravity from Strong Coupling Dynamics'', and by the MUR-PRIN2022 grant No.~2022NY2MXY.
M.B. is supported by MIUR PRIN Grant 2020KR4KN2 ``String Theory as a bridge between Gauge Theories and Quantum Gravity”. F.B. and M.B. are also supported by INFN Iniziativa Specifica ST\&FI. P.N. is supported by the Mani L. Bhaumik Institute for Theoretical Physics, by a DOE Early Career Award under DE-SC0020421, and by the Simons Collaboration on Global Categorical Symmetries.

\appendix

\section{Aspects of \tps{\matht{\mathfrak{su}(N)_k}}{su(N)k} Chern--Simons theory}
\label{app: su(N)k Chern-Simons}

Here we review some aspects of the three-dimensional gauge theories with algebra $\mathfrak{su}(N)$ and Chern--Simons level $k$. In particular, we first review the quantization of the level and the possible global forms of the gauge group, and then the corresponding 1-form and 0-form symmetries in each global variant.


\subsection{Level quantization and global variants}
\label{app: level quant and variants}

Let us first review the quantization condition on the Chern--Simons level $k$ in a three-dimensional theory with gauge group $SU(N)/\mathbb{Z}_p$, where $p$ is a divisor of $N$. We give two alternative derivations.

\paragraph{3d Chern--Simons term as 4d second Chern class.}
The Chern--Simons term on a 3d manifold $X_3$ can be defined as the instanton density integrated on a 4d manifold $\mathcal{M}_4$ with boundary, such that $\partial\mathcal{M}_4 = X_3$:
\be
S_\text{CS} = \frac{ik}{4\pi} \int_{X_3} \Tr \biggl( A \wedge dA - \frac{2i}{3} A \wedge A \wedge A \biggr) = \frac{2\pi i k}{8\pi^2} \int_{\mathcal{M}_4} \Tr (F \wedge F) \,.
\ee
In order to have a well-defined 3d Chern--Simons term, this expression should not depend on the choice of extension $\mathcal{M}_4$. Let us then consider two such extensions, $\mathcal{M}_4$ and $\mathcal{M}'_4$, and impose that the difference between the corresponding actions $S'_\text{CS} - S_\text{CS}$ takes values in $2\pi i \mathbb{Z}$, for every possible bundle. By gluing $\mathcal{M}'_4$ with $\overline{\mathcal{M}}_4$ (where bar denotes orientation reversal) along the common boundary $X_3$, this can be rewritten as the integral of the instanton density over the closed manifold $\mathcal{M}'_4 \cup \overline{\mathcal{M}}_4$. Using the relation between the quantization of the instanton number on closed manifolds and the magnetic flux (or second Stiefel-Whitney class, or Brauer class) $w_2 \in H^2(\mathcal{M}_4; \mathbb{Z}_p)$ of an $SU(N)/\mathbb{Z}_p$ bundle, namely (see \eg{} \cite{Hsin:2020nts})
\be
\frac1{8\pi^2} \int \Tr (F \wedge F) = \frac{N(p-1)}{2p^2} \int \mathcal{P}(w_2) \mod 1 \,,
\ee
where $\mathcal{P}$ is the Pontryagin square, one obtains the condition
\be
\label{k quantization instanton}
\frac{kN(p-1)}{2p^2} \int \mathcal{P}(w_2) \in \mathbb{Z} \,,
\ee
for every possible choice of $w_2$. 

On closed spin four-manifolds one has that $\int \mathcal{P}(w_2) \in 2\mathbb{Z}$, hence the condition \eqref{k quantization instanton} amounts to require that $k$ is a multiple of
\be
\label{k0 spin}
k_0^\text{spin}=\frac{p^2}{\gcd \bigl( p^2, N(p-1) \bigr)} = \frac{p}{\gcd \bigl(p,\frac{N}{p} \bigr)} \,.  
\ee
Note that $k_0^\text{spin} = 1$ (and hence any $k\in\mathbb{Z}$ is allowed) if and only if $p^2$ is a divisor of $N$ (this is trivially the case if $p=1$, leading to the $SU(N)_k$ CS theories, but also if $N = ap^2$ with $p>1$). On the other hand, if $p$ and $N/p$ are coprime then $k$ has to be a multiple of $p$ (this happens for example when $p=N$, namely in a $PSU(N)_k$ gauge theory). 

On non-spin manifolds one has that $\int \mathcal{P}(w_2) \in \mathbb{Z}$, hence the condition \eqref{k quantization instanton} implies that $k$ must be a multiple of
\be
\label{k0 bos}
k_0^\text{bos} = \frac{2p^2}{\gcd \bigl( 2p^2,N(p-1) \bigr)} = \frac{2p}{\gcd \bigl( 2p, \frac{N}{p}(p-1) \bigr)} =
\begin{cases} \dfrac{p}{ \gcd\bigl( p, \frac Np \bigr) } &\text{for $p$ odd} \,, \\[1em]
\dfrac{2p}{\gcd\bigl( 2p, \frac Np \bigr) } &\text{for $p$ even} \,. \end{cases}
\ee

\paragraph{Spin of 3d Wilson lines.}
Wilson lines charged under the $\mathbb{Z}_N$ 1-form symmetry of $SU(N)_k$ CS theory transform with a charge equal to the number of boxes in the Young diagram of the gauge representation used to define the line, \ie{} the $N$-ality of the representation. The topological line operators $U_m$ that implement the 1-form symmetry are themselves Wilson lines in pure Chern--Simons theory, in particular the symmetry defect $U_m$ is the Abelian Wilson line in representation $\mathcal{R}_{mk}$ whose Young diagram is a rectangle with $m$ rows and $k$ columns, with $m = 0,1,\ldots,N-1$. The spin of such Wilson lines can be read from the conformal dimensions $\Delta$ of the corresponding primary operators in the 2d $\mathfrak{su}(N)_k$ WZW model, which are given by
\be
\label{spin wilson lines}
\Delta(m) = \frac{C_2(\mathcal{R}_{mk})}{k+N} = \frac{km(N-m)}{2N} \,,
\ee
where $C_2(\mathcal{R}_{mk})$ is the quadratic Casimir of the representation $\mathcal{R}_{mk}$,%
\footnote{The quadratic Casimir (in the physicists' normalization) of the representation $\mathcal{R}_{mk}$ specified by the rectangular Young diagram with $m$ rows and $k$ columns can be obtained using the general formula:
\begin{equation*}
C_2(\mathcal{R}_{mk})= \frac{(\lambda,\lambda+2\rho)}{2} \,,   
\end{equation*}
where $\rho=[1,\dots,1]$ and $\lambda=[0,\dots,0,k,0,\dots,0]$ (where $k$ is at position $m$) are the $(N-1)$-dimensional $\su(N)$ Weyl vector and highest-weight vector of the representation $\mathcal{R}_{mk}$, respectively, in terms of Dynkin labels. The scalar product has matrix $G_{ij}=i(N-j)/N$ for $i\leq j$, and $G_{ij}=G_{ji}$ for $i>j$. We then get
\begin{equation*}
C_2(\mathcal{R}_{mk}) = \frac{1}{2} \sum_{i,j}^{N-1} G_{ij} \bigl( \delta_{im} \delta_{jm} k^2 + 2\delta_{im} k \bigr) = \frac{1}{2} \biggl( k^2 G_{mm} + 2k\sum_{j=1}^{N-1}G_{mj} \biggr) = \frac{km(N-m)(k+N)}{2N} \,.   
\end{equation*}}
while the spin is $\theta_m = e^{2\pi i \Delta(m)}$. We attempt to gauge a $\mathbb{Z}_p$ subgroup of the $\mathbb{Z}_N$ 1-form symmetry. Let $N=pq$, so that such a subgroup is implemented by the lines $U_{m'q}$ with $m' = 0,\ldots,p-1$. From \eqref{spin wilson lines}, the spin of these lines follows from
\be
\Delta(m'q) = \frac{kNm'(p-m')}{2p^2} \,.   
\ee
In order to be able to gauge $\bZ_p$ within spin theories, we need to impose $\Delta(m'q) \in \frac12 \bZ$ for all $m'$.
The most restrictive condition arises when $m'=1$, and it requires $k$ to be a multiple of $k_0^\text{spin}$ in (\ref{k0 spin}). In order to be able to gauge $\bZ_p$ within bosonic theories, we need to impose $\Delta(m'q) \in \bZ$ for all $m'$.
Once again, the most restrictive condition arises when $m'=1$ and it requires $k$ to be a multiple of $k_0^\text{bos}$ in (\ref{k0 bos}).


\subsection{Symmetries and anomalies}
\label{app: SU(N)_k/Z_p 1-form sym}

Let us discuss the 1-form symmetry of the global variants $\bigl( SU(N)/\bZ_p \bigr){}_k$ obtained by gauging a $\bZ_p$ subgroup of the $\bZ_N$ 1-form symmetry of $SU(N)_k$. Given the spin from \eqref{spin wilson lines} of the lines $U_m$ implementing the $\mathbb{Z}_N$ 1-form symmetry, their braiding is
\be
\label{braiding su(N)k}
\bB\left(m_1,m_2\right) = e^{2\pi i \left[ \rule{0pt}{0.6em} \Delta(m_1) + \Delta(m_2) - \Delta(m_1 + m_2) \right] } = \exp \biggl( 2\pi i \frac{k}N \, m_1 m_2 \biggr) \,.
\ee
The anomaly coefficient $\ell$ was defined in \cite{Hsin:2018vcg} by rewriting the spin \eqref{spin wilson lines} of the lines as
\be
\label{spin wilson lines with anomaly}
\theta(m) = \exp\biggl( - \pi i \ell \, \frac{m^2}{N} \biggr) \,, 
\ee
which shows that for bosonic theories $\ell$ is defined mod $2N$, and it is given by
\be
\label{anomaly coeff p=1}
\ell =
\begin{cases} k+N &\text{for $k$ odd} \,, \\
k &\text{for $k$ even} \,. \end{cases}
\ee
Notice that $N\ell$ is always even in bosonic theories. For spin theories, instead, the anomaly coefficient is defined mod $N$ and hence $\ell=k$ for all values of $k$.

We now proceed to gauge the lines $U_{m'q}$ implementing the $\mathbb{Z}_p$ subgroup of $\mathbb{Z}_N$, where $N=pq$.
Assuming that the quantization conditions on $k$ described in Appendix~\ref{app: level quant and variants} are met, this is an allowed operation. After gauging, the lines that remain as genuine symmetry defects are the ones which have a trivial braiding with the lines $U_{m'q}$, modded out by the identifications induced by fusion with $U_{m'q}$ \cite{Moore:1989yh}. The first requirement implies that the surviving symmetry defects $U_m$ are labeled by $m$'s such that $1 = \bB(m, m'q) = \exp \bigl( \frac{ 2\pi i k}{p} mm' \bigr)$ for all $m' \in \bZ$. This restricts $m=ra$, where
\be
a \,\equiv\, \frac{p}{\gcd(p,k)}
\ee
and $r$ is an integer. We then have to mod out this set by the action of fusion with the lines of $\bZ_p$, \ie{} we identify lines $U_m$ whose labels differ by multiples of $q$. We are thus left with the lines $U_{ra}$ labeled by $r = 0, \ldots, L-1$ which implement a $\bZ_L$ 1-form symmetry, where%
\footnote{Indeed the smallest identification between values of $m$ is by $\operatorname{lcm}(q,a) = qa/ \gcd(q,a)$. This means that the smallest identification between values of $r$ is by $L = \operatorname{lcm}(q,a)/a$.}
\be
L \,\equiv\, \frac{q}{\gcd(q, a)} \,.
\ee
As a consistency check, note that if $p=1$ then $L=N$ ($SU(N)_k$ has a $\mathbb{Z}_N$ 1-form symmetry), if $p=N$ then $L=1$ ($PSU(N)_k$ does not have a 1-form symmetry), and if $k=0$ then $L=q$ ($SU(N)/\mathbb{Z}_p$ has a $\mathbb{Z}_{N/p}$ 1-form symmetry).

Let us show that, for those $k$ that satisfy the quantization conditions \eqref{k0 spin} or \eqref{k0 bos}, the identity $\gcd(q,a) = a$ holds and therefore
\be
L = \frac qa = \frac{N \gcd(p,k)}{p^2} \;.
\ee
Indeed, if $k$ is a multiple of $p/\gcd(p,q)$, we can pull this factor out in the following $\gcd$:
\be
\gcd(p,k) = \frac{p}{\gcd(p,q)} \, \gcd\biggl( \gcd(p,q) \,,\, \frac{k}{p/\gcd(p,q)} \biggr) \,.
\ee
This allows us to write $a$ as
\be
a = \frac{ \gcd(p,q) }{ \gcd\bigl( \gcd(p,q) \,,\, \frac{k}{p/ \gcd(p,q)} \bigr) } \,,
\ee
which shows that $q$ is a multiple of $a$ and thus $\gcd(q,a) = a$ as sought. Similarly, if $k$ is a multiple of $2p/ \gcd(2p,q)$, we can pull this factor out and write 
\be
2 \gcd(p,k) = \gcd(2p, 2k) = \frac{2p}{\gcd(2p,q)} \, \gcd\biggl( \gcd(2p,q) \,,\, \frac{2k}{2p/ \gcd(2p,q)} \biggr) \,.
\ee
This allows us to write $a$ as
\be
a = \frac{\gcd(2p,q) }{ \gcd\bigl( \gcd(2p,q) \,,\, \frac{2k}{ 2p/\gcd(2p,q) } \bigr) } \,,
\ee
which again shows that $q$ is a multiple of $a$ and thus $\gcd(q,a) = a$ as sought.

Using \eqref{spin wilson lines}, the spin of the lines $U_{ra}$ that implement the $\bZ_L$ 1-form symmetry is given by
\be
\theta(ra) = \exp\biggl( \pi i \, \frac{k \, r \, ( L p - r ) }{ L \gcd(p,k) } \biggr) \,,
\ee
whereas their mutual braiding is
\be
\bB(r_1,r_2) = \exp \biggl( 2\pi i \, \frac{k}{L \gcd(p,k)} \, r_1 r_2 \biggr) \,.   
\ee
We can compute the anomaly coefficient $\ell_p$ of a given $\bigl( SU(N)/\mathbb{Z}_p \bigr){}_k$ global variant, which is now defined as $\theta(ra) = \exp\bigl( - \pi i \ell_p \frac{r^2}{L} \bigr)$. For bosonic theories $\ell_p$ is defined mod $2L$ and is given by
\be
\label{anomaly coeff p gen}
\ell_p =
\begin{cases} \dfrac{k}{\gcd(p,k)} + L &\text{for $k$ and $p$ odd} \,, \\[1em]
\dfrac{k}{\gcd(p,k)} &\text{otherwise} \,. \end{cases}
\ee
With some algebra and using (\ref{k0 bos}) one can show that $L \ell_p$ is always even.
For spin theories, instead, $\ell_p$ is defined mod $L$ and hence it is given by the latter case of the formula above for all values of $k$ and $p$. The result in \eqref{anomaly coeff p gen} matches the one in $\eqref{anomaly coeff p=1}$ for $p=1$.

Finally, let us discuss the magnetic 0-form symmetry in each global variant. When we gauge the $\bZ_p$ electric 1-form symmetry, a dual $\bZ_p$ magnetic 0-form symmetry arises. However, its $\bZ_{p/\gcd(p,k)}$ subgroup does not act on anything. Indeed, local operators with charge $m$ under the $\bZ_p$ 0-form symmetry are pointlike monopole operators, which can be identified as the endpoints of the lines $U_{mq}$ of the $\bZ_p$ 1-form symmetry we gauged. The only gauge-invariant monopole operators, though, are the ones with charge $m$ such that the corresponding $U_{mq}$ were trivial (endable) before gauging. This imposes $1 = \bB(m',mq) = \exp \bigl( \frac{ 2\pi i k}{p} mm' \bigr)$ for all $m' \in \bZ$, which restricts $m$ to be a multiple of $\frac{p}{\gcd(p,k)}$. Consequently, the $\bZ_{p/\gcd(p,k)}$ subgroup of the magnetic 0-form symmetry does not act on the gauge-invariant monopole operators, and only the quotient $\bZ_p / \bZ_{p/\gcd(p,k)} \cong \bZ_{\gcd(p,k)}$ acts faithfully.


\bibliographystyle{ytphys}
\baselineskip=0.87\baselineskip
\bibliography{SymTFT}

\providecommand{\href}[2]{#2}\begingroup\raggedright\begin{thebibliography}{10}

\bibitem{Gaiotto:2014kfa}
D.~Gaiotto, A.~Kapustin, N.~Seiberg, and B.~Willett, ``{Generalized Global
  Symmetries},'' \href{http://dx.doi.org/10.1007/JHEP02(2015)172}{{\em JHEP}
  {\bfseries 02} (2015) 172}, \href{http://arxiv.org/abs/1412.5148}{{\ttfamily
  arXiv:1412.5148 [hep-th]}}.

\bibitem{Wess:1971yu}
J.~Wess and B.~Zumino, ``{Consequences of anomalous Ward identities},''
  \href{http://dx.doi.org/10.1016/0370-2693(71)90582-X}{{\em Phys. Lett. B}
  {\bfseries 37} (1971) 95--97}.

\bibitem{Callan:1984sa}
C.~G. Callan, Jr. and J.~A. Harvey, ``{Anomalies and Fermion Zero Modes on
  Strings and Domain Walls},''
  \href{http://dx.doi.org/10.1016/0550-3213(85)90489-4}{{\em Nucl. Phys. B}
  {\bfseries 250} (1985) 427--436}.

\bibitem{Ji:2019eqo}
W.~Ji and X.-G. Wen, ``{Non-invertible anomalies and mapping-class-group
  transformation of anomalous partition functions},''
  \href{http://dx.doi.org/10.1103/PhysRevResearch.1.033054}{{\em Phys. Rev.
  Res.} {\bfseries 1} (2019) 033054},
  \href{http://arxiv.org/abs/1905.13279}{{\ttfamily arXiv:1905.13279
  [cond-mat.str-el]}}.

\bibitem{Gaiotto:2020iye}
D.~Gaiotto and J.~Kulp, ``{Orbifold groupoids},''
  \href{http://dx.doi.org/10.1007/JHEP02(2021)132}{{\em JHEP} {\bfseries 02}
  (2021) 132}, \href{http://arxiv.org/abs/2008.05960}{{\ttfamily
  arXiv:2008.05960 [hep-th]}}.

\bibitem{Apruzzi:2021nmk}
F.~Apruzzi, F.~Bonetti, I.~Garc{\'\i}a~Etxebarria, S.~S. Hosseini, and
  S.~Schafer-Nameki, ``{Symmetry TFTs from String Theory},''
  \href{http://dx.doi.org/10.1007/s00220-023-04737-2}{{\em Commun. Math. Phys.}
  {\bfseries 402} (2023) 895--949},
  \href{http://arxiv.org/abs/2112.02092}{{\ttfamily arXiv:2112.02092
  [hep-th]}}.

\bibitem{Freed:2022qnc}
D.~S. Freed, G.~W. Moore, and C.~Teleman, ``{Topological symmetry in quantum
  field theory},'' \href{http://arxiv.org/abs/2209.07471}{{\ttfamily
  arXiv:2209.07471 [hep-th]}}.

\bibitem{Freed:2014iua}
D.~S. Freed, ``{Anomalies and Invertible Field Theories},''
  \href{http://dx.doi.org/10.1090/pspum/088/01462}{{\em Proc. Symp. Pure Math.}
  {\bfseries 88} (2014) 25--46},
  \href{http://arxiv.org/abs/1404.7224}{{\ttfamily arXiv:1404.7224 [hep-th]}}.

\bibitem{Lin:2022dhv}
Y.-H. Lin, M.~Okada, S.~Seifnashri, and Y.~Tachikawa, ``{Asymptotic density of
  states in 2d CFTs with non-invertible symmetries},''
  \href{http://dx.doi.org/10.1007/JHEP03(2023)094}{{\em JHEP} {\bfseries 03}
  (2023) 094}, \href{http://arxiv.org/abs/2208.05495}{{\ttfamily
  arXiv:2208.05495 [hep-th]}}.

\bibitem{Bhardwaj:2023ayw}
L.~Bhardwaj and S.~Schafer-Nameki, ``{Generalized Charges, Part II:
  Non-Invertible Symmetries and the Symmetry TFT},''
  \href{http://arxiv.org/abs/2305.17159}{{\ttfamily arXiv:2305.17159
  [hep-th]}}.

\bibitem{Bartsch:2023wvv}
T.~Bartsch, M.~Bullimore, and A.~Grigoletto, ``{Representation theory for
  categorical symmetries},'' \href{http://arxiv.org/abs/2305.17165}{{\ttfamily
  arXiv:2305.17165 [hep-th]}}.

\bibitem{Kaidi:2023maf}
J.~Kaidi, E.~Nardoni, G.~Zafrir, and Y.~Zheng, ``{Symmetry TFTs and anomalies
  of non-invertible symmetries},''
  \href{http://dx.doi.org/10.1007/JHEP10(2023)053}{{\em JHEP} {\bfseries 10}
  (2023) 053}, \href{http://arxiv.org/abs/2301.07112}{{\ttfamily
  arXiv:2301.07112 [hep-th]}}.

\bibitem{Antinucci:2023ezl}
A.~Antinucci, F.~Benini, C.~Copetti, G.~Galati, and G.~Rizi, ``{Anomalies of
  non-invertible self-duality symmetries: fractionalization and gauging},''
  \href{http://arxiv.org/abs/2308.11707}{{\ttfamily arXiv:2308.11707
  [hep-th]}}.

\bibitem{Cordova:2023bja}
C.~C{\'o}rdova, P.-S. Hsin, and C.~Zhang, ``{Anomalies of Non-Invertible
  Symmetries in (3+1)d},'' \href{http://arxiv.org/abs/2308.11706}{{\ttfamily
  arXiv:2308.11706 [hep-th]}}.

\bibitem{Bhardwaj:2023fca}
L.~Bhardwaj, L.~E. Bottini, D.~Pajer, and S.~Schafer-Nameki, ``{Categorical
  Landau Paradigm for Gapped Phases},''
  \href{http://arxiv.org/abs/2310.03786}{{\ttfamily arXiv:2310.03786
  [cond-mat.str-el]}}.

\bibitem{Brennan:2024fgj}
T.~D. Brennan and Z.~Sun, ``{A SymTFT for Continuous Symmetries},''
  \href{http://arxiv.org/abs/2401.06128}{{\ttfamily arXiv:2401.06128
  [hep-th]}}.

\bibitem{Antinucci:2024zjp}
A.~Antinucci and F.~Benini, ``{Anomalies and gauging of $U(1)$ symmetries},''
  \href{http://arxiv.org/abs/2401.10165}{{\ttfamily arXiv:2401.10165
  [hep-th]}}.

\bibitem{Bonetti:2024cjk}
F.~Bonetti, M.~Del~Zotto, and R.~Minasian, ``{SymTFTs for Continuous
  non-Abelian Symmetries},'' \href{http://arxiv.org/abs/2402.12347}{{\ttfamily
  arXiv:2402.12347 [hep-th]}}.

\bibitem{Apruzzi:2024htg}
F.~Apruzzi, F.~Bedogna, and N.~Dondi, ``{SymTh for non-finite symmetries},''
  \href{http://arxiv.org/abs/2402.14813}{{\ttfamily arXiv:2402.14813
  [hep-th]}}.

\bibitem{Kaidi:2022cpf}
J.~Kaidi, K.~Ohmori, and Y.~Zheng, ``{Symmetry TFTs for Non-invertible
  Defects},'' \href{http://dx.doi.org/10.1007/s00220-023-04859-7}{{\em Commun.
  Math. Phys.} {\bfseries 404} (2023) 1021--1124},
  \href{http://arxiv.org/abs/2209.11062}{{\ttfamily arXiv:2209.11062
  [hep-th]}}.

\bibitem{Zhang:2023wlu}
C.~Zhang and C.~C\'ordova, ``{Anomalies of $(1+1)$-dimensional categorical
  symmetries},'' \href{http://dx.doi.org/10.1103/PhysRevB.110.035155}{{\em
  Phys. Rev. B} {\bfseries 110} (2024) 035155},
  \href{http://arxiv.org/abs/2304.01262}{{\ttfamily arXiv:2304.01262
  [cond-mat.str-el]}}.

\bibitem{Bhardwaj:2023wzd}
L.~Bhardwaj and S.~Schafer-Nameki, ``{Generalized Charges, Part I: Invertible
  Symmetries and Higher Representations},''
  \href{http://dx.doi.org/10.21468/SciPostPhys.16.4.093}{{\em SciPost Phys.}
  {\bfseries 16} (2024) 093}, \href{http://arxiv.org/abs/2304.02660}{{\ttfamily
  arXiv:2304.02660 [hep-th]}}.

\bibitem{Chen:2023qnv}
J.~Chen, W.~Cui, B.~Haghighat, and Y.-N. Wang, ``{SymTFTs and duality defects
  from 6d SCFTs on 4-manifolds},''
  \href{http://dx.doi.org/10.1007/JHEP11(2023)208}{{\em JHEP} {\bfseries 11}
  (2023) 208}, \href{http://arxiv.org/abs/2305.09734}{{\ttfamily
  arXiv:2305.09734 [hep-th]}}.

\bibitem{Bashmakov:2023kwo}
V.~Bashmakov, M.~Del~Zotto, and A.~Hasan, ``{Four-manifolds and Symmetry
  Categories of 2d CFTs},'' \href{http://arxiv.org/abs/2305.10422}{{\ttfamily
  arXiv:2305.10422 [hep-th]}}.

\bibitem{Sun:2023xxv}
Z.~Sun and Y.~Zheng, ``{When are Duality Defects Group-Theoretical?},''
  \href{http://arxiv.org/abs/2307.14428}{{\ttfamily arXiv:2307.14428
  [hep-th]}}.

\bibitem{Baume:2023kkf}
F.~Baume, J.~J. Heckman, M.~H\"ubner, E.~Torres, A.~P. Turner, and X.~Yu,
  ``{SymTrees and Multi-Sector QFTs},''
  \href{http://dx.doi.org/10.1103/PhysRevD.109.106013}{{\em Phys. Rev. D}
  {\bfseries 109} (2024) 106013},
  \href{http://arxiv.org/abs/2310.12980}{{\ttfamily arXiv:2310.12980
  [hep-th]}}.

\bibitem{Huang:2023pyk}
S.-J. Huang and M.~Cheng, ``{Topological holography, quantum criticality, and
  boundary states},'' \href{http://arxiv.org/abs/2310.16878}{{\ttfamily
  arXiv:2310.16878 [cond-mat.str-el]}}.

\bibitem{Bhardwaj:2023bbf}
L.~Bhardwaj, L.~E. Bottini, D.~Pajer, and S.~Schafer-Nameki, ``{The Club
  Sandwich: Gapless Phases and Phase Transitions with Non-Invertible
  Symmetries},'' \href{http://arxiv.org/abs/2312.17322}{{\ttfamily
  arXiv:2312.17322 [hep-th]}}.

\bibitem{Bhardwaj:2024qrf}
L.~Bhardwaj, D.~Pajer, S.~Schafer-Nameki, and A.~Warman, ``{Hasse Diagrams for
  Gapless SPT and SSB Phases with Non-Invertible Symmetries},''
  \href{http://arxiv.org/abs/2403.00905}{{\ttfamily arXiv:2403.00905
  [cond-mat.str-el]}}.

\bibitem{Hofman:2017vwr}
D.~M. Hofman and N.~Iqbal, ``{Generalized global symmetries and holography},''
  \href{http://dx.doi.org/10.21468/SciPostPhys.4.1.005}{{\em SciPost Phys.}
  {\bfseries 4} (2018) 005}, \href{http://arxiv.org/abs/1707.08577}{{\ttfamily
  arXiv:1707.08577 [hep-th]}}.

\bibitem{Bergman:2020ifi}
O.~Bergman, Y.~Tachikawa, and G.~Zafrir, ``{Generalized symmetries and
  holography in ABJM-type theories},''
  \href{http://dx.doi.org/10.1007/JHEP07(2020)077}{{\em JHEP} {\bfseries 07}
  (2020) 077}, \href{http://arxiv.org/abs/2004.05350}{{\ttfamily
  arXiv:2004.05350 [hep-th]}}.

\bibitem{Benini:2022hzx}
F.~Benini, C.~Copetti, and L.~Di~Pietro, ``{Factorization and global symmetries
  in holography},'' \href{http://dx.doi.org/10.21468/SciPostPhys.14.2.019}{{\em
  SciPost Phys.} {\bfseries 14} (2023) 019},
  \href{http://arxiv.org/abs/2203.09537}{{\ttfamily arXiv:2203.09537
  [hep-th]}}.

\bibitem{Apruzzi:2022rei}
F.~Apruzzi, I.~Bah, F.~Bonetti, and S.~Schafer-Nameki, ``{Noninvertible
  Symmetries from Holography and Branes},''
  \href{http://dx.doi.org/10.1103/PhysRevLett.130.121601}{{\em Phys. Rev.
  Lett.} {\bfseries 130} (2023) 121601},
  \href{http://arxiv.org/abs/2208.07373}{{\ttfamily arXiv:2208.07373
  [hep-th]}}.

\bibitem{Bergman:2022otk}
O.~Bergman and S.~Hirano, ``{The holography of duality in $\mathcal{N}{=}4$
  Super-Yang-Mills theory},''
  \href{http://dx.doi.org/10.1007/JHEP11(2022)069}{{\em JHEP} {\bfseries 11}
  (2022) 069}, \href{http://arxiv.org/abs/2208.09396}{{\ttfamily
  arXiv:2208.09396 [hep-th]}}.

\bibitem{Heckman:2022muc}
J.~J. Heckman, M.~H\"ubner, E.~Torres, and H.~Y. Zhang, ``{The Branes Behind
  Generalized Symmetry Operators},''
  \href{http://dx.doi.org/10.1002/prop.202200180}{{\em Fortsch. Phys.}
  {\bfseries 71} (2023) 2200180},
  \href{http://arxiv.org/abs/2209.03343}{{\ttfamily arXiv:2209.03343
  [hep-th]}}.

\bibitem{vanBeest:2022fss}
M.~van Beest, D.~S.~W. Gould, S.~Schafer-Nameki, and Y.-N. Wang, ``{Symmetry
  TFTs for 3d QFTs from M-theory},''
  \href{http://dx.doi.org/10.1007/JHEP02(2023)226}{{\em JHEP} {\bfseries 02}
  (2023) 226}, \href{http://arxiv.org/abs/2210.03703}{{\ttfamily
  arXiv:2210.03703 [hep-th]}}.

\bibitem{Antinucci:2022vyk}
A.~Antinucci, F.~Benini, C.~Copetti, G.~Galati, and G.~Rizi, ``{The holography
  of non-invertible self-duality symmetries},''
  \href{http://arxiv.org/abs/2210.09146}{{\ttfamily arXiv:2210.09146
  [hep-th]}}.

\bibitem{Bashmakov:2022uek}
V.~Bashmakov, M.~Del~Zotto, A.~Hasan, and J.~Kaidi, ``{Non-invertible
  symmetries of class $\mathcal{S}$ theories},''
  \href{http://dx.doi.org/10.1007/JHEP05(2023)225}{{\em JHEP} {\bfseries 05}
  (2023) 225}, \href{http://arxiv.org/abs/2211.05138}{{\ttfamily
  arXiv:2211.05138 [hep-th]}}.

\bibitem{Antinucci:2022cdi}
A.~Antinucci, C.~Copetti, G.~Galati, and G.~Rizi, ``{``Zoology'' of
  non-invertible duality defects: the view from class $\mathcal{S}$},''
  \href{http://dx.doi.org/10.1007/JHEP04(2024)036}{{\em JHEP} {\bfseries 04}
  (2024) 036}, \href{http://arxiv.org/abs/2212.09549}{{\ttfamily
  arXiv:2212.09549 [hep-th]}}.

\bibitem{Bah:2023ymy}
I.~Bah, E.~Leung, and T.~Waddleton, ``{Non-invertible symmetries, brane
  dynamics, and tachyon condensation},''
  \href{http://dx.doi.org/10.1007/JHEP01(2024)117}{{\em JHEP} {\bfseries 01}
  (2024) 117}, \href{http://arxiv.org/abs/2306.15783}{{\ttfamily
  arXiv:2306.15783 [hep-th]}}.

\bibitem{Apruzzi:2023uma}
F.~Apruzzi, F.~Bonetti, D.~S.~W. Gould, and S.~Schafer-Nameki, ``{Aspects of
  categorical symmetries from branes: SymTFTs and generalized charges},''
  \href{http://dx.doi.org/10.21468/SciPostPhys.17.1.025}{{\em SciPost Phys.}
  {\bfseries 17} (2024) 025}, \href{http://arxiv.org/abs/2306.16405}{{\ttfamily
  arXiv:2306.16405 [hep-th]}}.

\bibitem{Cvetic:2023pgm}
M.~Cveti\v{c}, J.~J. Heckman, M.~H\"ubner, and E.~Torres, ``{Generalized
  symmetries, gravity, and the swampland},''
  \href{http://dx.doi.org/10.1103/PhysRevD.109.026012}{{\em Phys. Rev. D}
  {\bfseries 109} (2024) 026012},
  \href{http://arxiv.org/abs/2307.13027}{{\ttfamily arXiv:2307.13027
  [hep-th]}}.

\bibitem{Yu:2023nyn}
X.~Yu, ``{Noninvertible symmetries in 2D from type IIB string theory},''
  \href{http://dx.doi.org/10.1103/PhysRevD.110.065008}{{\em Phys. Rev. D}
  {\bfseries 110} (2024) 065008},
  \href{http://arxiv.org/abs/2310.15339}{{\ttfamily arXiv:2310.15339
  [hep-th]}}.

\bibitem{Gould:2023wgl}
D.~S.~W. Gould, L.~Lin, and E.~Sabag, ``{Swampland constraints on the symmetry
  topological field theory of supergravity},''
  \href{http://dx.doi.org/10.1103/PhysRevD.109.126005}{{\em Phys. Rev. D}
  {\bfseries 109} (2024) 126005},
  \href{http://arxiv.org/abs/2312.02131}{{\ttfamily arXiv:2312.02131
  [hep-th]}}.

\bibitem{DelZotto:2024tae}
M.~Del~Zotto, S.~N. Meynet, and R.~Moscrop, ``{Remarks on geometric
  engineering, symmetry TFTs and anomalies},''
  \href{http://dx.doi.org/10.1007/JHEP07(2024)220}{{\em JHEP} {\bfseries 07}
  (2024) 220}, \href{http://arxiv.org/abs/2402.18646}{{\ttfamily
  arXiv:2402.18646 [hep-th]}}.

\bibitem{etingof2016tensor}
P.~Etingof, S.~Gelaki, D.~Nikshych, and V.~Ostrik,
  \href{http://dx.doi.org/10.1090/surv/205}{{\em {Tensor categories}}},
  vol.~205 of {\em Mathematical Surveys and Monographs}.
\newblock American Math. Soc., 2015.
\newblock \url{https://math.mit.edu/~etingof/tenscat.pdf}.

\bibitem{turaev1992state}
V.~G. Turaev and O.~Y. Viro, ``{State sum invariants of 3-manifolds and quantum
  $6j$-symbols},'' \href{http://dx.doi.org/10.1016/0040-9383(92)90015-A}{{\em
  Topology} {\bfseries 31} (1992) 865--902}.

\bibitem{Thorngren:2019iar}
R.~Thorngren and Y.~Wang, ``{Fusion category symmetry. Part I. Anomaly in-flow
  and gapped phases},'' \href{http://dx.doi.org/10.1007/JHEP04(2024)132}{{\em
  JHEP} {\bfseries 04} (2024) 132},
  \href{http://arxiv.org/abs/1912.02817}{{\ttfamily arXiv:1912.02817
  [hep-th]}}.

\bibitem{Crane:1993if}
L.~Crane and D.~Yetter,
  \href{http://dx.doi.org/10.1142/9789812796387_0005}{``{A categorical
  construction of 4D topological quantum field theories},''} in {\em {Quantum
  topology}}, L.~H. Kauffman and R.~A. Baadhio, eds., pp.~120--130.
\newblock World Scientific, 1993.
\newblock \href{http://arxiv.org/abs/hep-th/9301062}{{\ttfamily
  arXiv:hep-th/9301062}}.

\bibitem{Crane:1994ji}
L.~Crane, L.~H. Kauffman, and D.~N. Yetter, ``{State sum invariants of
  four-manifolds},'' \href{http://dx.doi.org/10.1142/S0218216597000145}{{\em J.
  Knot Th. and Its Ramifications} {\bfseries 6} (1997) 177--234},
  \href{http://arxiv.org/abs/hep-th/9409167}{{\ttfamily arXiv:hep-th/9409167}}.

\bibitem{Levin:2004mi}
M.~A. Levin and X.-G. Wen, ``{String net condensation: A physical mechanism for
  topological phases},''
  \href{http://dx.doi.org/10.1103/PhysRevB.71.045110}{{\em Phys. Rev. B}
  {\bfseries 71} (2005) 045110},
  \href{http://arxiv.org/abs/cond-mat/0404617}{{\ttfamily
  arXiv:cond-mat/0404617}}.

\bibitem{douglas2018fusion}
C.~L. Douglas and D.~J. Reutter, ``{Fusion 2-categories and a state-sum
  invariant for 4-manifolds},''
  \href{http://arxiv.org/abs/1812.11933}{{\ttfamily arXiv:1812.11933
  [math.QA]}}.

\bibitem{Barkeshli:2016mew}
M.~Barkeshli, P.~Bonderson, C.-M. Jian, M.~Cheng, and K.~Walker, ``{Reflection
  and time reversal symmetry enriched topological phases of matter: path
  integrals, non-orientable manifolds, and anomalies},''
  \href{http://dx.doi.org/10.1007/s00220-019-03475-8}{{\em Commun. Math. Phys.}
  {\bfseries 374} (2019) 1021--1124},
  \href{http://arxiv.org/abs/1612.07792}{{\ttfamily arXiv:1612.07792
  [cond-mat.str-el]}}.

\bibitem{Bulmash:2020flp}
D.~Bulmash and M.~Barkeshli, ``{Absolute anomalies in (2+1)D symmetry-enriched
  topological states and exact (3+1)D constructions},''
  \href{http://dx.doi.org/10.1103/PhysRevResearch.2.043033}{{\em Phys. Rev.
  Res.} {\bfseries 2} (2020) 043033},
  \href{http://arxiv.org/abs/2003.11553}{{\ttfamily arXiv:2003.11553
  [cond-mat.str-el]}}.

\bibitem{Tata:2021jwp}
S.~Tata, R.~Kobayashi, D.~Bulmash, and M.~Barkeshli, ``{Anomalies in (2+1)D
  Fermionic Topological Phases and (3+1)D Path Integral State Sums for
  Fermionic SPTs},'' \href{http://dx.doi.org/10.1007/s00220-022-04484-w}{{\em
  Commun. Math. Phys.} {\bfseries 397} (2023) 199--336},
  \href{http://arxiv.org/abs/2104.14567}{{\ttfamily arXiv:2104.14567
  [cond-mat.str-el]}}.

\bibitem{Walker:2021esr}
K.~Walker, ``{A universal state sum},''
  \href{http://arxiv.org/abs/2104.02101}{{\ttfamily arXiv:2104.02101
  [math.QA]}}.

\bibitem{Barkeshli:2014cna}
M.~Barkeshli, P.~Bonderson, M.~Cheng, and Z.~Wang, ``{Symmetry
  Fractionalization, Defects, and Gauging of Topological Phases},''
  \href{http://dx.doi.org/10.1103/PhysRevB.100.115147}{{\em Phys. Rev. B}
  {\bfseries 100} (2019) 115147},
  \href{http://arxiv.org/abs/1410.4540}{{\ttfamily arXiv:1410.4540
  [cond-mat.str-el]}}.

\bibitem{Witten:1998wy}
E.~Witten, ``{AdS / CFT correspondence and topological field theory},''
  \href{http://dx.doi.org/10.1088/1126-6708/1998/12/012}{{\em JHEP} {\bfseries
  12} (1998) 012}, \href{http://arxiv.org/abs/hep-th/9812012}{{\ttfamily
  arXiv:hep-th/9812012}}.

\bibitem{Heckman:2024oot}
J.~J. Heckman, M.~H\"ubner, and C.~Murdia, ``{On the holographic dual of a
  topological symmetry operator},''
  \href{http://dx.doi.org/10.1103/PhysRevD.110.046007}{{\em Phys. Rev. D}
  {\bfseries 110} (2024) 046007},
  \href{http://arxiv.org/abs/2401.09538}{{\ttfamily arXiv:2401.09538
  [hep-th]}}.

\bibitem{Witten:1998zw}
E.~Witten, ``{Anti-de Sitter space, thermal phase transition, and confinement
  in gauge theories},''
  \href{http://dx.doi.org/10.4310/ATMP.1998.v2.n3.a3}{{\em Adv. Theor. Math.
  Phys.} {\bfseries 2} (1998) 505--532},
  \href{http://arxiv.org/abs/hep-th/9803131}{{\ttfamily arXiv:hep-th/9803131}}.

\bibitem{Fujita:2009kw}
M.~Fujita, W.~Li, S.~Ryu, and T.~Takayanagi, ``{Fractional Quantum Hall Effect
  via Holography: Chern-Simons, Edge States, and Hierarchy},''
  \href{http://dx.doi.org/10.1088/1126-6708/2009/06/066}{{\em JHEP} {\bfseries
  06} (2009) 066}, \href{http://arxiv.org/abs/0901.0924}{{\ttfamily
  arXiv:0901.0924 [hep-th]}}.

\bibitem{Rey:2008zz}
S.-J. Rey, ``{String theory on thin semiconductors: Holographic realization of
  Fermi points and surfaces},''
  \href{http://dx.doi.org/10.1143/PTPS.177.128}{{\em Prog. Theor. Phys. Suppl.}
  {\bfseries 177} (2009) 128--142},
  \href{http://arxiv.org/abs/0911.5295}{{\ttfamily arXiv:0911.5295 [hep-th]}}.

\bibitem{Hong:2010sb}
D.~K. Hong and H.-U. Yee, ``{Holographic aspects of three dimensional QCD from
  string theory},'' \href{http://dx.doi.org/10.1007/JHEP05(2010)036}{{\em JHEP}
  {\bfseries 05} (2010) 036}, \href{http://arxiv.org/abs/1003.1306}{{\ttfamily
  arXiv:1003.1306 [hep-th]}}. [Erratum: JHEP 08, 120 (2010)].

\bibitem{Argurio:2020her}
R.~Argurio, A.~Armoni, M.~Bertolini, F.~Mignosa, and P.~Niro, ``{Vacuum
  structure of large $N$ QCD$_{3}$ from holography},''
  \href{http://dx.doi.org/10.1007/JHEP07(2020)134}{{\em JHEP} {\bfseries 07}
  (2020) 134}, \href{http://arxiv.org/abs/2006.01755}{{\ttfamily
  arXiv:2006.01755 [hep-th]}}.

\bibitem{Argurio:2022vtz}
R.~Argurio and A.~Caddeo, ``{Comments on holographic level/rank dualities},''
  \href{http://dx.doi.org/10.1007/JHEP08(2022)097}{{\em JHEP} {\bfseries 08}
  (2022) 097}, \href{http://arxiv.org/abs/2205.06115}{{\ttfamily
  arXiv:2205.06115 [hep-th]}}.

\bibitem{Kitaev:2005hzj}
A.~Kitaev, ``{Anyons in an exactly solved model and beyond},''
  \href{http://dx.doi.org/10.1016/j.aop.2005.10.005}{{\em Annals Phys.}
  {\bfseries 321} (2006) 2--111},
  \href{http://arxiv.org/abs/cond-mat/0506438}{{\ttfamily
  arXiv:cond-mat/0506438}}.

\bibitem{Hsin:2018vcg}
P.-S. Hsin, H.~T. Lam, and N.~Seiberg, ``{Comments on One-Form Global
  Symmetries and Their Gauging in 3d and 4d},''
  \href{http://dx.doi.org/10.21468/SciPostPhys.6.3.039}{{\em SciPost Phys.}
  {\bfseries 6} (2019) 039}, \href{http://arxiv.org/abs/1812.04716}{{\ttfamily
  arXiv:1812.04716 [hep-th]}}.

\bibitem{Moore:1989yh}
G.~W. Moore and N.~Seiberg, ``{Taming the Conformal Zoo},''
  \href{http://dx.doi.org/10.1016/0370-2693(89)90897-6}{{\em Phys. Lett. B}
  {\bfseries 220} (1989) 422--430}.

\bibitem{Dijkgraaf:1989pz}
R.~Dijkgraaf and E.~Witten, ``{Topological Gauge Theories and Group
  Cohomology},'' \href{http://dx.doi.org/10.1007/BF02096988}{{\em Commun. Math.
  Phys.} {\bfseries 129} (1990) 393}.

\bibitem{Maldacena:2001ss}
J.~M. Maldacena, G.~W. Moore, and N.~Seiberg, ``{D-brane charges in five-brane
  backgrounds},'' \href{http://dx.doi.org/10.1088/1126-6708/2001/10/005}{{\em
  JHEP} {\bfseries 10} (2001) 005},
  \href{http://arxiv.org/abs/hep-th/0108152}{{\ttfamily arXiv:hep-th/0108152}}.

\bibitem{Banks:2010zn}
T.~Banks and N.~Seiberg, ``{Symmetries and Strings in Field Theory and
  Gravity},'' \href{http://dx.doi.org/10.1103/PhysRevD.83.084019}{{\em Phys.
  Rev. D} {\bfseries 83} (2011) 084019},
  \href{http://arxiv.org/abs/1011.5120}{{\ttfamily arXiv:1011.5120 [hep-th]}}.

\bibitem{Kapustin:2014gua}
A.~Kapustin and N.~Seiberg, ``{Coupling a QFT to a TQFT and Duality},''
  \href{http://dx.doi.org/10.1007/JHEP04(2014)001}{{\em JHEP} {\bfseries 04}
  (2014) 001}, \href{http://arxiv.org/abs/1401.0740}{{\ttfamily arXiv:1401.0740
  [hep-th]}}.

\bibitem{Putrov:2016qdo}
P.~Putrov, J.~Wang, and S.-T. Yau, ``{Braiding Statistics and Link Invariants
  of Bosonic/Fermionic Topological Quantum Matter in 2+1 and 3+1 dimensions},''
  \href{http://dx.doi.org/10.1016/j.aop.2017.06.019}{{\em Annals Phys.}
  {\bfseries 384} (2017) 254--287},
  \href{http://arxiv.org/abs/1612.09298}{{\ttfamily arXiv:1612.09298
  [cond-mat.str-el]}}.

\bibitem{Ye:2014oua}
P.~Ye and Z.-C. Gu, ``{Vortex-Line Condensation in Three Dimensions: A Physical
  Mechanism for Bosonic Topological Insulators},''
  \href{http://dx.doi.org/10.1103/PhysRevX.5.021029}{{\em Phys. Rev. X}
  {\bfseries 5} (2015) 021029},
  \href{http://arxiv.org/abs/1410.2594}{{\ttfamily arXiv:1410.2594
  [cond-mat.str-el]}}.

\bibitem{Zhang:2023ynd}
Z.-F. Zhang, Q.-R. Wang, and P.~Ye, ``{Continuum field theory of
  three-dimensional topological orders with emergent fermions and braiding
  statistics},'' \href{http://dx.doi.org/10.1103/PhysRevResearch.5.043111}{{\em
  Phys. Rev. Res.} {\bfseries 5} (2023) 043111},
  \href{http://arxiv.org/abs/2307.09983}{{\ttfamily arXiv:2307.09983
  [cond-mat.str-el]}}.

\bibitem{Fujita:2016gmu}
M.~Fujita, C.~M. Melby-Thompson, R.~Meyer, and S.~Sugimoto, ``{Holographic
  Chern-Simons Defects},''
  \href{http://dx.doi.org/10.1007/JHEP06(2016)163}{{\em JHEP} {\bfseries 06}
  (2016) 163}, \href{http://arxiv.org/abs/1601.00525}{{\ttfamily
  arXiv:1601.00525 [hep-th]}}.

\bibitem{Maldacena:1998im}
J.~M. Maldacena, ``{Wilson loops in large $N$ field theories},''
  \href{http://dx.doi.org/10.1103/PhysRevLett.80.4859}{{\em Phys. Rev. Lett.}
  {\bfseries 80} (1998) 4859--4862},
  \href{http://arxiv.org/abs/hep-th/9803002}{{\ttfamily arXiv:hep-th/9803002}}.

\bibitem{Rey:1998ik}
S.-J. Rey and J.-T. Yee, ``{Macroscopic strings as heavy quarks in large $N$
  gauge theory and anti-de Sitter supergravity},''
  \href{http://dx.doi.org/10.1007/s100520100799}{{\em Eur. Phys. J. C}
  {\bfseries 22} (2001) 379--394},
  \href{http://arxiv.org/abs/hep-th/9803001}{{\ttfamily arXiv:hep-th/9803001}}.

\bibitem{Witten:1979ey}
E.~Witten, ``{Dyons of Charge $e \theta / 2 \pi$},''
  \href{http://dx.doi.org/10.1016/0370-2693(79)90838-4}{{\em Phys. Lett. B}
  {\bfseries 86} (1979) 283--287}.

\bibitem{noi}
R.~Argurio, F.~Benini, M.~Bertolini, G.~Galati, and P.~Niro.
\newblock Work in progress.

\bibitem{Witten:1988hf}
E.~Witten, ``{Quantum Field Theory and the Jones Polynomial},''
  \href{http://dx.doi.org/10.1007/BF01217730}{{\em Commun. Math. Phys.}
  {\bfseries 121} (1989) 351--399}.

\bibitem{Drukker:1999zq}
N.~Drukker, D.~J. Gross, and H.~Ooguri, ``{Wilson loops and minimal
  surfaces},'' \href{http://dx.doi.org/10.1103/PhysRevD.60.125006}{{\em Phys.
  Rev. D} {\bfseries 60} (1999) 125006},
  \href{http://arxiv.org/abs/hep-th/9904191}{{\ttfamily arXiv:hep-th/9904191}}.

\bibitem{Drukker:2005kx}
N.~Drukker and B.~Fiol, ``{All-genus calculation of Wilson loops using
  D-branes},'' \href{http://dx.doi.org/10.1088/1126-6708/2005/02/010}{{\em
  JHEP} {\bfseries 02} (2005) 010},
  \href{http://arxiv.org/abs/hep-th/0501109}{{\ttfamily arXiv:hep-th/0501109}}.

\bibitem{Yamaguchi:2006tq}
S.~Yamaguchi, ``{Wilson loops of anti-symmetric representation and
  D5-branes},'' \href{http://dx.doi.org/10.1088/1126-6708/2006/05/037}{{\em
  JHEP} {\bfseries 05} (2006) 037},
  \href{http://arxiv.org/abs/hep-th/0603208}{{\ttfamily arXiv:hep-th/0603208}}.

\bibitem{Gomis:2006sb}
J.~Gomis and F.~Passerini, ``{Holographic Wilson Loops},''
  \href{http://dx.doi.org/10.1088/1126-6708/2006/08/074}{{\em JHEP} {\bfseries
  08} (2006) 074}, \href{http://arxiv.org/abs/hep-th/0604007}{{\ttfamily
  arXiv:hep-th/0604007}}.

\bibitem{Gomis:2006im}
J.~Gomis and F.~Passerini, ``{Wilson Loops as D3-Branes},''
  \href{http://dx.doi.org/10.1088/1126-6708/2007/01/097}{{\em JHEP} {\bfseries
  01} (2007) 097}, \href{http://arxiv.org/abs/hep-th/0612022}{{\ttfamily
  arXiv:hep-th/0612022}}.

\bibitem{Hsin:2020nts}
P.-S. Hsin and H.~T. Lam, ``{Discrete theta angles, symmetries and
  anomalies},'' \href{http://dx.doi.org/10.21468/SciPostPhys.10.2.032}{{\em
  SciPost Phys.} {\bfseries 10} (2021) 032},
  \href{http://arxiv.org/abs/2007.05915}{{\ttfamily arXiv:2007.05915
  [hep-th]}}.

\end{thebibliography}\endgroup

\end{document}